\documentclass[12pt]{article}
\newcommand {\citeAY} [1] {\citeNP {#1}}%
\newcommand {\citeAPY}[1] {\citeN  {#1}}%
\usepackage{xchicago}
\renewcommand {\showoriginalref}[1]{}
\renewcommand {\showCODEN}[1]{}
\renewcommand {\showISSN}[1]{}
\renewcommand {\showMR}[3]{}

\newcommand\eq[1] {(\ref{#1})}

\newcommand{\bfm}[1]{\mbox{\boldmath ${#1}$}}
\newcommand{\nonum}{\nonumber \\}
\newcommand{\beqa}{\begin{eqnarray}}
\newcommand{\eeqa}[1]{\label{#1}\end{eqnarray}}
\newcommand{\beq}{\begin{equation}}
\newcommand{\eeq}[1]{\label{#1}\end{equation}}
\newcommand{\Grad}{\nabla}
\newcommand{\Div}{\nabla \cdot}
\newcommand{\Curl}{\nabla \times}

\newcommand{\lang}{\langle}
\newcommand{\rang}{\rangle}
\newcommand{\Md}{\partial}

\newcommand{\Ga}{\alpha}

\newcommand{\Gd}{\delta}

\newcommand{\Gve}{\varepsilon}

\newcommand{\Gk}{\kappa}

\newcommand{\Gl}{\lambda}

\newcommand{\Gm}{\mu}

\newcommand{\Gt}{\theta}

\newcommand{\Gr}{\rho}

\newcommand{\Gs}{\sigma}

\newcommand{\Gj}{\tau}

\newcommand{\Go}{\omega}

\newcommand{\GG}{\Gamma}

\newcommand{\GO}{\Omega}

\newcommand{\GY}{\Psi}

\newcommand{\BGe}{\bfm\epsilon}
\newcommand{\BGve}{\bfm\varepsilon}

\newcommand{\BGm}{\bfm\mu}

\newcommand{\BGr}{\bfm\rho}

\newcommand{\BGs}{\bfm\sigma}

\newcommand{\BGj}{\bfm\tau}

\newcommand{\CA}{{\cal A}}

\newcommand{\CF}{{\cal F}}
\newcommand{\CG}{{\cal G}}

\newcommand{\CL}{{\cal L}}

\newcommand{\CT}{{\cal T}}

\newcommand{\BCC}{{\bfm{\cal C}}}

\newcommand{\BCE}{{\bfm{\cal E}}}

\newcommand{\BCK}{{\bfm{\cal K}}}
\newcommand{\BCL}{{\bfm{\cal L}}}
\newcommand{\BCM}{{\bfm{\cal M}}}
\newcommand{\BCN}{{\bfm{\cal N}}}

\newcommand{\BCP}{{\bfm{\cal P}}}

\newcommand{\BCR}{{\bfm{\cal R}}}

\def\Bb{{\bf b}}

\def\Be{{\bf e}}
\def\Bf{{\bf f}}
\def\Bg{{\bf g}}
\def\Bh{{\bf h}}

\def\Bj{{\bf j}}

\def\Bm{{\bf m}}
\def\Bn{{\bf n}}

\def\Bp{{\bf p}}
\def\Bq{{\bf q}}
\def\Br{{\bf r}}
\def\Bs{{\bf s}}
\def\Bt{{\bf t}}
\def\Bu{{\bf u}}
\def\Bv{{\bf v}}

\def\Bx{{\bf x}}

\def\BA{{\bf A}}
\def\BB{{\bf B}}
\def\BC{{\bf C}}
\def\BD{{\bf D}}
\def\BE{{\bf E}}
\def\BF{{\bf F}}
\def\BG{{\bf G}}
\def\BH{{\bf H}}
\def\BI{{\bf I}}
\def\BJ{{\bf J}}
\def\BK{{\bf K}}
\def\BL{{\bf L}}
\def\BM{{\bf M}}
\def\BN{{\bf N}}

\def\BT{{\bf T}}

\def\BZ{{\bf Z}}

\def \ba {\begin{array}}
\def \ea {\end{array}}
\newtheorem {Thm} {Theorem} [section]

\newtheorem {Adef} [Thm] {Definition}

\newtheorem {Arem} [Thm] {Remark}

\newtheorem {Aexa} [Thm] {Example}

\newtheorem {Anot} [Thm] {Notation}

\def \refe #1.{(\ref{#1})}
\def \reff #1.{figure~\ref{#1}}
\def \refs #1.{section~\ref{#1}}
\def \refss #1.{subsection~\ref{#1}}
\def \refD #1.{Definition~\ref{#1}}
\def \refT #1.{Theorem~\ref{#1}}
\def \refL #1.{Lemma~\ref{#1}}
\def \refC #1.{Corollary~\ref{#1}}
\def \refP #1.{Proposition~\ref{#1}}
\def \refR #1.{Remark~\ref{#1}}
\def \refE #1.{Example~\ref{#1}}
\def \refN #1.{Notation~\ref{#1}}

\usepackage{graphics}
\usepackage{amssymb}
\begin{document}   
\vspace{-1in}
\title{Minimization variational principles for acoustics, elastodynamics, and electromagnetism in lossy 
inhomogeneous bodies at fixed frequency}
\author{Graeme W. Milton \\
\small{Department of Mathematics, University of Utah, Salt Lake City UT 84112, USA}\\ {~~}\\
Pierre Seppecher and Guy Bouchitt\'e\\
\small{Institut de Math\'ematiques de Toulon},\\
\small{Universit\'e de Toulon et du Var, BP 132-83957 La Garde Cedex, France}}
\date{}
\maketitle
\begin{abstract}
The classical energy minimization principles of Dirichlet and Thompson are extended
as minimization principles
to acoustics, elastodynamics and electromagnetism in lossy inhomogeneous bodies
at fixed frequency. This is done by building upon ideas of Cherkaev and Gibiansky, who derived
minimization variational principles for quasistatics. In the
absence of free current the primary electromagnetic minimization variational principles
have a minimum which is the time-averaged electrical power dissipated in the
body. The variational principles provide constraints on the boundary values of the
fields when the moduli are known. Conversely, when the boundary values of the fields
have been measured, then they provide information about the values of the moduli
within the body. This should have application to electromagnetic tomography. 
We also derive saddle point variational principles which correspond to variational
principles of Gurtin, Willis, and Borcea. 
 
\end{abstract}
\vskip2mm

\noindent Keywords: variational principles, acoustics, elastodynamics, electromagnetism
\section{Introduction}
As our goal is to extend the classical minimization principles to acoustics, elastodynamics,
and electromagnetism, let us begin by recalling the well-known classical results
of electrostatics. Consider the equation of electrostatics
\beq \Div\Gve\Grad V=0, \eeq{0.1}
for the real valued potential $V(\Bx)$ in a body $\GG$ 
containing a locally isotropic material
with scalar real positive dielectric constant $\Gve(\Bx)$. This equation can
be rewritten in the equivalent form 
\beq \BD=\Gve\BE,\quad \BE=-\Grad V, \quad \Div\BD=0,
\eeq{0.1a}
where the first relation is known as the constitutive relation 
linking the electric displacement field $\BD(\Bx)$ and the electrical
field $\BE(\Bx)$ where these two fields satisfy the differential constraints
embodied in the second two relations. Two variational principles are
well known. We have the Dirichlet variational principle that
\beq  W(V)=\inf_{\underline{V}}W(\underline{V})\quad {\rm where}~
W(\underline{V})=\int_{\GG}\Gve |\Grad \underline{V}|^2/2,
\eeq{0.2}
in which the infimum is taken over all trial potentials $\underline{V}(\Bx)$ satisfying the
boundary condition that $\underline{V}(\Bx)=V(\Bx)$ on the boundary $\Md\GG$.
We also have dual Thompson variational principle that 
\beq \widetilde{W}(\BD)=\inf_{\matrix{\underline{\BD} \cr \Div\underline{\BD}=0}}
\widetilde{W}(\underline{\BD})\quad {\rm where}~
\widetilde{W}(\underline{\BD})=\int_{\GG}
|\underline{\BD}|^2/(2\Gve),
\eeq{0.2a}
where the infimum is taken over all divergence free trial displacement fields
$\underline{\BD}$ satisfying the boundary condition that 
$\underline{\BD}\cdot\Bn=\BD\cdot\Bn$ on  $\Md\GG$. 
The quantity $\Gve |\Grad V|^2/2=|\BD|^2/(2\Gve)$
is the electrostatic energy density inside the body, and so these
variational principles are also known as energy minimization principles.
In the mathematically analogous DC electrical conductivity problem,
where $\Gve(\Bx)$ and $\BD(\Bx)$ are replaced by the conductivity $\Gs(\Bx)$ 
and the current $\BJ(\Bx)$, this quantity is half the electrical power
dissipation and in that context the variational principles could
more accurately be called power dissipation minimization principles.

An appealing feature of these variational principles is that 
the minimum values $W(V)$ and $\widetilde{W}(\BD)$
can be expressed in terms of the values 
of $V(\Bx)$ and $\BD\cdot\Bn$ at the boundary $\Md\GG$:
\beq W(V)=\widetilde{W}(\BD)=-\frac{1}{2}\int_{\Md\GG}(\BD\cdot\Bn)V(\Bx).
\eeq{0.3}
So these variational principles provide a constraint on the
values of $V_0(\Bx)=V(\Bx)$ and $q_0(\Bx)\equiv -\BD(\Bx)\cdot\Bn$ at the boundary $\Md\GG$.
The relationship between these boundary functions $q_0$ and $V_0$ is linear
and the relationship $q_0=\BN V_0$ defines the Dirichlet to Neumann
map $\BN$, which is a real self-adjoint positive semi-definite operator.
Therefore $\BN$ is determined by knowledge of the quadratic form
\beq w(V_0)=\langle V_0,\BN V_0\rangle/2,
\eeq{0.4}
where for any two scalar or vector, real or complex, valued functions  $\Bp(\Bx)$ and
$\Bq(\Bx)$ defined on the boundary  $\Md\GG$, we denote
\beq \lang \Bp,\Bq\rang=\int_{\Md\GG}\Bp(\Bx)\cdot\Bq(\Bx).
\eeq{0.5} 
Given trial potentials $\underline{V}(\Bx, V_0)$,
parameterized by the function $V_0$,
with $\underline{V}(\Bx, V_0)=V_0(\Bx)$ on $\Md\GG$, the Dirichlet variational 
principle implies the inequality
\beq w(V_0)\leq W(\underline{V}), \eeq{0.6}
and provides an ``upper bound'' on the map $\BN$. By ``upper bound'' we mean an
upper bound on the associated non-negative quadratic form $w(V_0)$. 

To obtain a ``lower bound'' one
uses the Thompson variational principle. Since $w(V_0)=0$ when $V_0$ is constant
the null-space of $\BN$ consists of all constant fields and so 
$\BN$ is determined by knowledge of the quadratic form
\beq \widetilde{w}(q_0)=\langle q_0,\BN^{-1} q_0\rangle/2,
\eeq{0.7}
for all fields $q_0(\Bx)$ with zero average value on $\Md\GG$. The functionals
$w(V_0)$ and $\widetilde{w}(q_0)$ are Legendre transforms of each other,
\beq w(V_0)=\inf_{q}[\lang V_0,q\rang-\widetilde{w}(q)], \eeq{0.8}
where the infimum is over all fields $q(\Bx)$ with zero average value on $\Md\GG$,
and so an upper bound on the functional $\widetilde{w}(q)$ provides a lower
bound on $w(V_0)$. Given trial displacement fields $\underline{\BD}(\Bx, q_0)$
parameterized by the function $q_0$,
with $-\underline{\BD}(\Bx)\cdot\Bn=q_0(\Bx)$ on $\Md\GG$, the Thompson variational 
principle implies the inequality 
\beq \widetilde{w}(q_0)\leq \widetilde{W}(\underline{\BD}),
\eeq{0.9}
and provides a ``lower bound'' on the map $\BN$. 

Thus the Dirichlet and Thompson variational principles allow one to bound
the Dirichlet to Neuman map $\BN$. As shown by \citeAPY{Berryman:1990:VCE}
they can also be used in an inverse fashion:
if $\BN$ and hence the functionals $w(V_0)$ and $\widetilde{w}(q_0)$ are
known one can obtain information about the permittivity distribution
$\Gve(\Bx)$ from the inequalities \eq{0.6} and \eq{0.9} by appropriately
chosing the trial potentials  $\underline{V}(\Bx, V_0)$ and trial
displacement fields $\underline{\BD}(\Bx, q_0)$. It has been proved
that knowledge of $\BN$ uniquely determines $\Gve(\Bx)$ when 
$\Gve(\Bx)$ is piecewise analytic (\citeAY{Kohn:1984:DCB}),
or more generally when
$\Gve(\Bx)$ is smooth (\citeAY{Sylvester:1987:GUT}).

In this paper we extend these variational principles to 
acoustics, elastodynamics, and electromagnetism, in lossy
inhomogeneous bodies at fixed frequency. By lossy we mean that the time-averaged dissipation
of mechanical or electrical energy into heat is positive everywhere,
which means that the imaginary part of certain tensors, such 
as the permittivity tensor, are positive definite. Like the Dirichlet and Thompson
variational principles, the variational principles we obtain are minimization
(not saddle point) variational principles. Other variational
principles which extend the Dirichlet and Thompson variational principles to dynamics, not assuming time 
harmonicity, were derived for elastodynamics and electromagnetism by
\citeAPY{Gurtin:1964:VPL} and Willis (\citeyearNP{Willis:1981:VPDP}, \citeyearNP{Willis:1984:VPOE}).
These are minimizing variational principles in the Laplace domain but not in the
frequency domain. We will recover the Gurtin-Willis variational principles in the time
harmonic setting where they correspond to saddle point variational principles.
We remark that \citeAPY{Ben-Amoz:1966:VPA} and
Willis (\citeyearNP{Willis:1981:VPDP}, \citeyearNP{Willis:1984:VPOE}) 
derived variational principles which are dynamic analogs of the 
Hashin-Shtrikman (\citeyearNP{Hashin:1962:VAT}, \citeyearNP{Hashin:1962:SVP}, 
\citeyearNP{Hashin:1962:VATE}, \citeyearNP{Hashin:1963:VAT})
variational principles, involving trial polarizations fields. We 
do not consider such variational principles here.

Of course given a (inhomogeneous) linear differential operator $\CA$,
mapping $m$-component complex vector fields to $m$-component complex vector fields,
there is a trivial minimization variational principle
associated with the equation  $\CA\Bu=\Bh$ for the $m$-component 
potential $\Bu$, in which $\Bh$ is a $m$-component source term.
When the equation $\CA\Bu=\Bh$ with $\Bu=\Bu_0$ on $\Md\GG$ has
a unique solution for $\Bu$, the infimum in 
\beq 0=\inf_{\underline{\Bu}}\int_{\GG} 
\overline{(\CA\underline{\Bu}-\Bh)} \cdot(\CA\underline{\Bu}-\Bh),
\eeq{0.9a}
where the overline denotes complex conjugation and
the infimum is over all fields $\underline{\Bu}$ with 
$\underline{\Bu}=\Bu_0$ on $\Md\GG$, is clearly attained 
when $\underline{\Bu}=\Bu$. What separates such variational
principles from those of Dirichlet and Thompson is that, as explained
above, the latter variational principles give us useful information on the boundary
values of the fields. We desire variational principles with this property. Also, from
a mathematical viewpoint, the existance of a solution cannot be deduced from 
\eq{0.9a}, and in case there is a solution no interesting inequality results
from \eq{0.9a}.

In an important development \citeAPY{Cherkaev:1994:VPC} 
extended the Dirichlet and Thompson variational principles
to the quasistatic electromagnetic equations 
(and also to the quasistatic elastodynamic equations)
where the equations \eq{0.1} and \eq{0.1a}
still hold but $\Gve(\Bx)$, $V(\Bx)$, $\BE(\Bx)$ and $\BD(\Bx)$ are complex, and the imaginary
part of $\Gve(\Bx)$ is positive. Since we build upon their work let us briefly review
their ideas in this context.

By rewriting the constitutive law $\BD=\Gve\BE$ in terms of
its real and imaginary parts 
\beq \pmatrix{\BD'' \cr \BD'}=\pmatrix{\Gve'' & \Gve' \cr \Gve' & -\Gve''}
\pmatrix{\BE' \cr \BE''},
\eeq{0.10}
where the prime denotes the real part, while the double prime denotes the imaginary part, 
it is evident that since $\Gve''(\Bx)>0$ one has the Cherkaev-Gibiansky
saddle point variational principle 
\beq Q(V',V'')=\inf_{\underline{V}'}\sup_{\underline{V}''}Q(\underline{V}',\underline{V}'')
=\sup_{\underline{V}''}\inf_{\underline{V}'}Q(\underline{V}',\underline{V}''),
\eeq{0.11}
where
\beq Q(\underline{V}',\underline{V}'')=
\int_{GG}\pmatrix{\Grad \underline{V}' \cr \Grad \underline{V}''}\cdot 
\pmatrix{\Gve'' & \Gve' \cr \Gve' & -\Gve''}\pmatrix{\Grad \underline{V}' \cr \Grad \underline{V}''},
\eeq{0.12}
and the infimum and supremum are over trial potentials  
with $\underline{V}'(\Bx)=V'(\Bx)$ and  $\underline{V}''(\Bx)=V''(\Bx)$
on the boundary $\Md\GG$.
To obtain a minimization variational principle from this saddle point variational principle
\citeAPY{Cherkaev:1994:VPC} make a partial Legendre transform of the 
saddle shaped quadratic form  
to convert it into a convex quadratic form. This is
equivalent to rewriting the constitutive law in the form
\beq  \pmatrix{\BD'' \cr \BE''}= \BCE\pmatrix{\BE' \cr -\BD'},
\eeq{0.13}
where the matrix 
\beq \BCE  =  \pmatrix{\Gve''+(\Gve')^2/(\Gve'') & \Gve'/\Gve'' \cr 
                   \Gve'/\Gve'' & 1/\Gve''} 
\eeq{0.14}
is positive definite when $\Gve''(\Bx)$ is positive. From this it is evident that one has the
Cherkaev-Gibiansky minimization variational principle:
\beq Y(V',\BD')=\inf_{\underline{V}'}
\inf_{\matrix{\underline{\BD}' \cr \Div\underline{\BD}'=0}}Y(\underline{V}',\underline{\BD}'),
\eeq{0.15}
where 
\beq Y(\underline{V}',\underline{\BD}')
=\int_\GG\pmatrix{\Grad \underline{V}' \cr \underline{\BD}'}\cdot\BCE\pmatrix{\Grad \underline{V}' \cr \underline{\BD}'},
\eeq{0.16}
and the infimums are over all trial real potentials with $\underline{V}'(\Bx)=V'(\Bx)$ on  $\Md\GG$, and all
trial real divergence free displacement fields with $\underline{\BD}\cdot\Bn=\BD\cdot\Bn$ on  $\Md\GG$.

As an aside we remark that, as shown in section 18 of \citeAPY{Milton:1990:CSP}, 
one can extend these ideas of Gibiansky and Cherkaev
to obtain variational principles for problems where the constitutive law takes the form $\BD=\BGve\BE$, where
the matrix valued field $\BGve(\Bx)$ is not symmetric. For example, for conduction in a
fixed magnetic field, the conductivity tensor $\BGs(\Bx)$ is real but not symmetric. Expressing 
$\BGs(\Bx)=\BGs_s(\Bx)+\BGs_a(\Bx)$, where $\BGs_s(\Bx)=\BGs_s^T(\Bx)$ and
$\BGs_a(\Bx)=-\BGs_a^T(\Bx)$ are the symmetric and antisymmetric parts of the conductivity tensor,
one considers current
fields $\BJ(\Bx)$ and electric fields $\BE(\Bx)$ which solve the conductivity equations
\beq \BJ=(\BGs_s+\BGs_a)\BE, \quad\BE=-\Grad V, \quad \Div\BJ=0,
\eeq{0.17}
in conjunction with current
fields $\widehat{\BJ}(\Bx)$ and electric fields $\widehat{\BE}(\Bx)$ which solve the adjoint problem
\beq \widehat{\BJ}=(\BGs_s-\BGs_a)\widehat{\BE}, \quad\widehat{\BE}=-\Grad \widehat{V}, \quad \Div\widehat{\BJ}=0.
\eeq{0.18}
By adding and subtracting the two constitutive laws one gets a new constitutive law
\beq \pmatrix{\BJ_s \cr -\BJ_a}=\pmatrix{\BGs_s &  \BGs_a \cr -\BGs_a & -\BGs_s}\pmatrix{\BE_s \cr \BE_a},
\eeq{0.19}
involving fields
\beq \BJ_s\equiv\BJ+\widehat{\BJ}, \quad \BJ_a\equiv\BJ-\widehat{\BJ}, \quad
\BE_s\equiv\BE+\widehat{\BE}, \quad \BE_a\equiv\BE-\widehat{\BE},
\eeq{0.20}
that satisfy the differential constraints
\beq \Div\BJ_s=\Div\BJ_a=0, \quad {\BE}_s=-\Grad {V}_s,\quad {\BE}_a=-\Grad{V}_a,
\eeq{0.21}
where $V_s\equiv V+\widehat{V}$ and $V_a\equiv V-\widehat{V}$. Conversely, given any fields which solve
\eq{0.19} and \eq{0.21}, one can recover the associated fields which solve \eq{0.17} and \eq{0.18}. 
The two problems are equivalent. Since 
the matrix entering the constitutive law \eq{0.19} is real and symmetric there is clearly an associated 
saddle point variational principle when $\BGs_s(\Bx)$ is positive definite for all $\Bx$. \citeAPY{Fannjiang:1994:CED} show how this variational principle provides useful asymptotic estimates for
the mathematically analogous problem of convection enhanced diffusion. Also by
rewriting the constitutive law \eq{0.19} in the equivalent form
\beq \pmatrix{\BJ_s \cr \BE_a}
=\pmatrix{\BGs_s-\BGs_a\BGs_s^{-1}\BGs_a &  \BGs_a\BGs_s^{-1} \cr -\BGs_s^{-1}\BGs_a & \BGs_s^{-1}}\pmatrix{\BE_s \cr \BJ_a},
\eeq{0.22}
which involves a matrix which is symmetric and positive definite when $\BGs_s(\Bx)$ is positive definite for all $\Bx$,
one obtains an associated minimization variational principle (\citeAY{Milton:1990:CSP}).
In this paper we do not treat such cases where
the constitutive laws involve non-symmetric matrices: this will be covered elsewhere. 

\section{General Theory}
\setcounter{equation}{0}
Let $d$ be the dimension of the space and let
$\CF(\Bx)$ and $\CG(\Bx)$ be two complex valued 
fields of the form
\beq \CF=\pmatrix{\BF \cr \Bf},\quad \CG=\pmatrix{\BG \cr \Bg},
\eeq{1.1}
where $\BF(\Bx)$ and $\BG(\Bx)$ are $m\times d$ dimensional
matrix fields and $\Bf(\Bx)$ and $\Bg(\Bx)$ 
are $m$ component vector fields,
and which satisfy the differential constraints
\beq \CF=\sqcap \Bu,\quad \Bh+\sqcup\CG=0,\eeq{1.2}
where $\Bu(\Bx)$ is an $m$-component potential and
$\Bh(\Bx)$ is some $m$-component source term and 
\beq \sqcap u\equiv\pmatrix{\Grad \Bu \cr \Bu},\quad 
\sqcup\CG\equiv -\Div\BG+\Bg
\eeq{1.3}
serve to define the operators $\sqcap$ and $\sqcup$. The differential
constraints imply the key property that on any domain $\GG$
\beq \int_{\GG}\CG\cdot\CF+\Bh\cdot\Bu~
=\int_{\Md\GG}(\BG\cdot\Bn)\cdot\Bu, 
\eeq{1.4}
as follows from integration by parts, where 
\beq \CG\cdot\CF\equiv \BG\cdot\BF+\Bg\cdot\Bf,
\quad \BG\cdot\BF\equiv{\rm Tr}(\BG^T\BF),
\eeq{1.5}
and $\Bn$ is the outward unit normal to the surface $\Md\GG$. 
(Note that \eq{1.5} is an inner product only when $\CG$ and $\CF$
are real, since it does not involve complex conjugation). Here, 
following standard notation, we define
\beq \{\Grad \Bu\}_{ij}=\frac{\Md u_i}{\Md x_j}, \quad
 \{\Div\BG\}_i=\sum_{j=1}^d \frac{\Md G_{ij}}{\Md x_j}.
\eeq{1.5a}
Now suppose the fields
$\CF$ and $\CG$ are linked by the constitutive relation
\beq \CG(\Bx)=\BZ(\Bx)\CF(\Bx), \eeq{1.6}
where $\BZ(\Bx)=\BZ'(\Bx)+i\BZ''(\Bx)$ is a complex symmetric 
linear map (which is represented by a complex symmetric 
$m(d+1)\times m(d+1)$ matrix when
$\CG$ and $\CF$ are represented by $m(d+1)$ component vectors).
The map $\BZ$ can be expressed in block form as
\beq \BZ=\pmatrix{\BL & \BK \cr \BK^T & \BM}, \eeq{1.6aa}
where (for each $\Bx$) 
$\BL(\Bx)$ is a symmetric linear map on the space of $m\times d$ dimensional matrices,
$\BK(\Bx)$ is a linear map from $m$ component vectors to 
$m\times d$ dimensional matrices, $\BK^T(\Bx)$ is a linear map from 
$m\times d$ dimensional matrices to $m$ component vectors, and
$\BM(\Bx)$ is a symmetric linear map on the space of $m$ component vectors.
Thus, written in terms of components, \eq{1.6} implies
\beqa G_{ij}& = & L_{ijk\ell}F_{k\ell}+K_{ijk}f_{k}, \nonum
g_{i} &= & K_{k\ell i}F_{k\ell}+M_{ik}f_{k},
\eeqa{1.6ab}
in which sums of repeated indices are assumed. Since $\BL$ and $\BM$
are symmetric maps, we have
\beq L_{ijk\ell}=L_{k\ell ij},\quad M_{ik}=M_{ki}. \eeq{1.6c}

The constitutive law and differential constraints imply
that $\Bu$ satisfies
\beq \Bh+\sqcup(\BZ\sqcap \Bu)=0,
\eeq{1.7} 
or equivalently,
\beq \Div(\BL\Grad\Bu+\BK\Bu)=\Bh+\BK^T\Grad\Bu+\BM\Bu.
\eeq{1.7a}
Our analysis thus applies to equations of this form, with 
$\BL$ and $\BM$ having the symmetries \eq{1.6c}.

Let $(\Bu,\CF,\CG)$ be a solution of \eq{1.2} and \eq{1.6}.
Taking real and imaginary parts, the differential constraints imply
\beq \CF'=\sqcap \Bu',\quad \Bh'+\sqcup\CG'=0,\quad
 \CF''=\sqcap \Bu'',\quad \Bh''+\sqcup\CG''=0,
\eeq{1.8}
where the prime denotes the real part and the double prime the imaginary 
part, and the constitutive law can be written in the
form 
\beq \pmatrix{\CG'' \cr \CG'}
=\pmatrix{\BZ'' & \BZ' \cr \BZ' & -\BZ''}\pmatrix{\CF' \cr \CF''}.
\eeq{1.9}

To make use of this representation let us further suppose that 
the imaginary part of $\BZ$ satisfies
\beq \BZ''(\Bx)\geq \Ga\BI,\quad\forall \Bx\in\GG,
\eeq{1.9a}   
for some $\Ga>0$. 
Extending the ideas of \citeAPY{Cherkaev:1994:VPC} consider the real 
valued quadratic functional
\beqa Q({\underline\Bu}',{\underline\Bu}'') & = & \int_\GG 
\pmatrix{\sqcap \underline{\Bu}' \cr \sqcap {\underline\Bu}''}\cdot\pmatrix{\BZ'' & \BZ' \cr \BZ' & -\BZ''}
\pmatrix{\sqcap {\underline\Bu}' \cr \sqcap {\underline\Bu}''}+2(\Bh''\cdot{\underline\Bu}'+\Bh'\cdot{\underline\Bu}'') \nonum
& = & \int_\GG[(\sqcap\underline{\Bu})\cdot\BZ\sqcap\underline{\Bu}+2\Bh\cdot\Bu]'',
\eeqa{1.10}
subject to the boundary conditions  that ${\underline\Bu}'=\Bu'$ and ${\underline\Bu}''=\Bu''$ on 
$\Md\GG$. Let $\Bs(\Bx)$ be a real valued $m$ component vector field
with $\Bs=0$ on $\Md\GG$. Now
$Q({\underline\Bu}'+\Bs,{\underline\Bu}'')$ is a convex quadratic function of $\Bs$
(with a linear component) 
while $Q({\underline\Bu}',{\underline\Bu}''+\Bs)$ is a concave
quadratic function of $\Bs$ (with a linear component).
Therefore on the space of fields satisfying the boundary conditions
the pair $({\underline\Bu}',{\underline\Bu}'')$
which are at the saddle point of the variational problem
\beq \inf_{{\underline\Bu}'}\sup_{{\underline\Bu}''}Q({\underline\Bu}',{\underline\Bu}''), \eeq{1.11}
exists and is unique and corresponds to the stationary point of $Q({\underline\Bu}',{\underline\Bu}'')$,
and is also at the saddle point of the variational problem
\beq \sup_{{\underline\Bu}''}\inf_{{\underline\Bu}'}Q({\underline\Bu}',{\underline\Bu}''). \eeq{1.11a}

Now let us prove we have the variational principle
\beq Q(\Bu',\Bu'')=\inf_{{\underline\Bu}'}\sup_{{\underline\Bu}''}Q({\underline\Bu}',{\underline\Bu}'') 
=\sup_{{\underline\Bu}''}\inf_{{\underline\Bu}'}Q({\underline\Bu}',{\underline\Bu}''),
\eeq{1.14aa}
where the infimum and supremum are over fields with ${\underline\Bu}'$ and ${\underline\Bu}''$ 
with any fixed values $\Bu_0'$ and $\Bu_0''$
at the boundary $\Md\GG$ and $\Bu'$ and $\Bu''$ are the solutions
with these prescribed boundary conditions.

Let $(\Bu,\CF,\CG)$ be a solution of \eq{1.2} and \eq{1.6}.
As $\Bs=0$ on $\Md\GG$, in a similar way as \eq{1.4} was derived, we get
\beq \int_{\GG}\CG''\cdot(\sqcap \Bs)+\Bh''\cdot\Bs=0, \quad
 \int_{\GG}\CG'\cdot(\sqcap \Bs)+\Bh'\cdot\Bs=0.
\eeq{1.12}
Thus we see that
\beqa Q(\Bu'+\Bs,\Bu'')& = & Q(\Bu',\Bu'')+
\int_{\GG}2\CG''\cdot(\sqcap \Bs)+2\Bh''\cdot\Bs
+(\sqcap \Bs)\cdot\BZ''\sqcap \Bs \nonum
&= & Q(\Bu',\Bu'')+\int_{\GG}(\sqcap \Bs)\cdot\BZ''\sqcap \Bs,
\eeqa{1.13}
and in the same way we get
\beq Q(\Bu',\Bu''+\Bs)=  Q(\Bu',\Bu'')-\int_{\GG}(\sqcap \Bs)\cdot\BZ''\sqcap \Bs.
\eeq{1.14}
So we see that the saddle point of \eq{1.11} is at $({\underline\Bu}',{\underline\Bu}'')=(\Bu',\Bu'')$. 

Conversely, let $(\Bu',\Bu'')$ be a saddle point of \eq{1.11} satisfying some boundary 
condition ${\Bu}'=\Bu_0'$ and ${\Bu}''=\Bu_0''$ on $\Md\GG$.
Consider
\beq \inf_{\Bs}\sup_{\Bt}Q(\Bu'+\Bs,\Bu''+\Bt), 
\eeq{1.14a}
where the infimum and supremum are taken over fields $\Bs$ and $\Bt$ satisfying $\Bs=\Bt=0$ on $\Md\GG$. 
A necessary condition for the saddle point to be
at $\Bs=\Bt=0$ is that the first-order derivative of the functional $Q$ vanishes: that is
\beq \int_\GG 
2\pmatrix{\sqcap \Bs \cr \sqcap \Bt }\cdot
\pmatrix{{\CG}'' \cr {\CG}'}+2\Bh''\cdot\Bs+2\Bh'\cdot\Bt=0,
\eeq{1.14b}
for any $\Bs$ and $\Bt$ satisfying $\Bs=\Bt=0$ on $\Md\GG$, where we have introduced
\beq 
\pmatrix{{\CG}'' \cr {\CG}'}\equiv
\pmatrix{\BZ'' & \BZ' \cr \BZ' & -\BZ''}\cdot
\pmatrix{\sqcap {\Bu}' \cr \sqcap {\Bu}''}.
\eeq{1.14c}
Upon setting
\beq {\CG}'=\pmatrix{{\BG}' \cr {\Bg}'},\quad 
{\CG}''=\pmatrix{{\BG}'' \cr {\Bg}''},
\eeq{1.14d}
the condition \eq{1.14b} reduces to
\beq \int_\GG{\BG}''\cdot\Grad\Bs+({\Bg}''+\Bh'')\cdot\Bs
+{\BG}'\cdot\Grad\Bt+({\Bg}'+\Bh')\cdot\Bt
=0.
\eeq{1.14e}
Integrating this by parts gives
\beq \int_\GG(-\Div{\BG}''+{\Bg}''+\Bh'')\cdot\Bs
+(-\Div{\BG}'+{\Bg}'+\Bh')\cdot\Bt =0,
\eeq{1.14f}
and this will be satisfied for any $\Bs$ and $\Bt$ satisfying $\Bs=\Bt=0$ 
on $\Md\GG$ if and only if
\beq \quad \Bh'+\sqcup{\CG}'=0,\quad \Bh''+\sqcup{\CG}''=0.
\eeq{1.14g}
Thus the Euler-Lagrange equations associated with the saddle point variational
principle coincide with the original equations \eq{1.2}. 

There is also a dual saddle point variational principle, analogous to the Thompson
variational principle. By taking real and imaginary parts of
the constitutive relation $\CF=\BK\CG$ where $\BK=\BZ^{-1}$ has negative 
definite imaginary part we obtain 
\beq  \pmatrix{\CF'' \cr \CF'}
=\pmatrix{\BK'' & \BK' \cr \BK' & -\BK''}\pmatrix{\CG' \cr \CG''}.
\eeq{1.14h}
The dual variational principle involves the quadratic form 
\beq R({\underline\BG}',{\underline\BG}'')=\int_\GG  
\pmatrix{{\underline\CG}' \cr {\underline\CG}''}\cdot\pmatrix{\BK'' & \BK' \cr \BK' & -\BK''}\pmatrix{{\underline\CG}' \cr {\underline\CG}''}
=\int_\GG(\underline\CG\cdot\BK\underline\CG)'',
\eeq{1.14i}
where ${\underline\CG}'$ and ${\underline\CG}''$ are given by
\beq {\underline\CG}'=\pmatrix{{\underline\BG}' \cr \Div{\underline\BG}'-\Bh'},\quad
{\underline\CG}''=\pmatrix{{\underline\BG}'' \cr \Div{\underline\BG}''-\Bh''},
\eeq{1.14j}
and ${\underline\BG}'$ and ${\underline\BG}''$ satisfy the boundary conditions
\beq {\underline\BG}'\cdot\Bn=\BG'\cdot\Bn
\quad {\rm and~} {\underline\BG}''\cdot\Bn=\BG''\cdot\Bn \quad{\rm on~} \Md\GG, 
\eeq{1.14k}
where $\BG'$ and $\BG''$ are the real and imaginary parts of $\BG$. Notice that
the form of ${\underline\CG}$ given by \eq{1.14j} ensures that $\Bh+\sqcup{\underline\CG}=0$. 
inside $\GG$. By similar argument to
the one establishing \eq{1.14aa}, we have the dual variational principle
\beq R(\BG',\BG'')=\sup_{{\underline\BG}'}\inf_{{\underline\BG}''}R({\underline\BG}',{\underline\BG}'')
=\inf_{{\underline\BG}''}\sup_{{\underline\BG}'}R({\underline\BG}',{\underline\BG}''),
\eeq{1.14l}
where the supremum and infimum are over fields satisfying \eq{1.14k}.
  
Again building upon the ideas of \citeAPY{Cherkaev:1994:VPC} let us rewrite 
the constitutive relation in the form
\beq \pmatrix{\CG'' \cr \CF''}=\BCL\pmatrix{\CF' \cr -\CG'}, 
\eeq{1.15}
where straightforward algebra shows that 
\beq \BCL=\pmatrix{\BZ''+\BZ'(\BZ'')^{-1}\BZ' & \BZ'(\BZ'')^{-1} \cr 
                   (\BZ'')^{-1}\BZ' & (\BZ'')^{-1}}.
\eeq{1.16}
For all $\Bx\in\GG$ the matrix $\BCL(\Bx)$ is positive definite
since, using \eq{1.9} and \eq{1.15}, we see that the associated quadratic form
\beqa \pmatrix{\CF' \cr -\CG'}\cdot\BCL\pmatrix{\CF' \cr -\CG'} & = &
\CF'\cdot\CG''-\CF''\cdot\CG' \nonum
&=& \CF'\cdot\BZ''\CF'+\CF''\cdot\BZ''\CF''
\eeqa{1.17}
is non-negative, and zero only when $\CF'=\CF''=0$, i.e. when
$\CF'=\CG'=0$.  

Now consider the convex functional
\beq Y({\underline\Bu}',{\underline\BG}')=\int_\GG 
\pmatrix{\sqcap {\underline\Bu}' \cr -{\underline\CG}'}\cdot\BCL
\pmatrix{\sqcap {\underline\Bu}' \cr -{\underline\CG}'}+2\Bh''\cdot{\underline\Bu}',
\eeq{1.18}
where ${\underline\CG}'$ is given by
\beq {\underline\CG}'=\pmatrix{{\underline\BG}' \cr \Div{\underline\BG}'-\Bh'}, \eeq{1.18aa}
(to ensure that $\Bh'+\sqcup{\underline\CG}'=0$ inside $\GG$)  
and ${\underline\Bu}'$ and ${\underline\BG}'$ are real and satisfy the boundary conditions that 
\beq {\underline\Bu}'=\Bu'_0,\quad {\rm and~} {\underline\BG}'\cdot\Bn=\BG'_0\cdot\Bn \quad{\rm on~} \Md\GG.
\eeq{1.18a}
Let us prove we have the variational principle
\beq Y(\Bu',\BG')=\inf_{{\underline\Bu}'}\inf_{{\underline\BG}'}Y({\underline\Bu}',{\underline\BG}'), \eeq{1.20}
where the infimums are over all trial fields ${\underline\Bu}'$ and ${\underline\BG}'$
satisfying the boundary conditions \eq{1.18a}. Let $(\Bu,\CG)$ be a solution of \eq{1.2} and \eq{1.6}
with ${\Bu}'=\Bu'_0$ and ${\BG}'\cdot\Bn=\BG'_0\cdot\Bn$ on $\Md\GG$. Suppose we are given any 
$m$ component vector field $\Bs(\Bx)$ with $\Bs=0$ on $\Md\GG$ and also a 
$m\times d$ matrix valued field $\BT(\Bx)$ with $\BT\cdot\Bn=0$ on $\Md\GG$.
Then \eq{1.12} and integration by parts implies that
\beq \int_{\GG}2\CG''\cdot(\sqcap \Bs)+2\Bh''\cdot\Bs-2\CT\cdot(\sqcap \Bu'')=0,\quad
{\rm where}~\CT=\pmatrix{\BT \cr \Div\BT}. \eeq{1.20a}
From this identity it follows that
\beq
Y(\Bu'+\Bs,\BG'+\BT) =  Y(\Bu',\BG')+
\int_{\GG}\pmatrix{\sqcap \Bs \cr -\CT}\cdot\BCL\pmatrix{\sqcap \Bs \cr -\CT},
\eeq{1.21}
which implies $(\Bu',\BG')$ is a mimimizer of the variational principle.

Conversely, let $(\Bu',\BG')$ be a mimimizer of the variational principle, satisfying some
boundary conditions  ${\Bu}'=\Bu'_0$ and ${\BG}'\cdot\Bn=\BG'_0\cdot\Bn$ on $\Md\GG$. Consider
\beq \inf_{\Bs}\inf_{\BT}Y({\Bu}'+\Bs,{\BG}'+\BT),
\eeq{1.21a}
where the infimum is over all fields $\Bs$ and $\BT$ satisfying $\Bs=0$ 
and $\BT\cdot\Bn=0$ on $\Md\GG$. A necessary condition for the infimum to be
at $\Bs=\BT=0$ is that the first-order derivative of the functional $Y$ vanish: that is
\beq \int_\GG 
2\pmatrix{\sqcap \Bs \cr -\CT}\cdot
\pmatrix{{\CG}'' \cr {\CF}''}+2\Bh''\cdot{\Bs}=0,
\eeq{1.21b}
for any $\Bs$ and $\BT$ satisfying $\Bs=0$ 
and $\BT\cdot\Bn=0$ on $\Md\GG$, where we have introduced
\beq \CT\equiv\pmatrix{\BT \cr \Div\BT}, \quad 
\pmatrix{{\CG}'' \cr {\CF}''}\equiv
\BCL\pmatrix{\sqcap {\Bu}' \cr -{\CG}'}.
\eeq{1.21c}
Upon setting
\beq {\CF}''=\pmatrix{{\BF}'' \cr {\Bf}''},\quad 
{\CG}''=\pmatrix{{\BG}'' \cr {\Bg}''},
\eeq{1.21d}
the condition \eq{1.21b} reduces to
\beq \int_\GG{\BG}''\cdot\Grad\Bs+({\Bg}''+\Bh'')\cdot\Bs
-\BT\cdot{\BF}''-(\Div\BT)\cdot{\Bf}''=0.
\eeq{1.21e}
Integrating this by parts gives
\beq \int_\GG(-\Div{\BG}''+{\Bg}''+\Bh'')\cdot\Bs
-\BT\cdot({\BF}''-\Grad{\Bf}'')=0,
\eeq{1.21f}
and this will be satisfied for any $\Bs$ and $\BT$ satisfying $\Bs=0$ 
and $\BT\cdot\Bn=0$ on $\Md\GG$ if and only if
\beq \Bh''+\sqcup{\CG}''=0, \quad {\CF}''=\sqcap{\Bf}''.
\eeq{1.21g}
Upon identifying ${\Bf}''$ with $\Bu''$
we see that the Euler-Lagrange equations associated with the minimization variational
principle coincide with the original equations \eq{1.2}. 

The convex variational principle we have just stated is, as usual, associated with a
dual variational principle:
\beq \widetilde{Y}(\Bu'',\BG'')=\inf_{{\underline\Bu}''}\inf_{{\underline\BG}''}
\widetilde{Y}({\underline\Bu}'',{\underline\BG}''),
\eeq{1.21n}
where
\beq \widetilde{Y}({\underline\Bu}'',{\underline\BG}'')=\int_\GG 
\pmatrix{{\underline\CG}''\cr \sqcap {\underline\Bu}'' }\cdot\BCL^{-1}
\pmatrix{{\underline\CG}''\cr \sqcap {\underline\Bu}'' }-2\Bh'\cdot{\underline\Bu}'',
\eeq{1.21o}
in which
\beq {\underline\CG}''=\pmatrix{{\underline\BG}'' \cr \Div{\underline\BG}''-\Bh''},
\eeq{1.21p}
and the infimum in \eq{1.21n} is over real fields ${\underline\Bu}''$ and ${\underline\BG}''$ with
${\underline\Bu}''=\Bu''$ and ${\underline\BG}''\cdot\Bn=\BG''\cdot\Bn$ on $\Md\GG$.
Since 
\beq \BCL^{-1}=\pmatrix{(\BZ'')^{-1} & - (\BZ'')^{-1}\BZ' \cr 
                       -\BZ'(\BZ'')^{-1} & \BZ''+\BZ'(\BZ'')^{-1}\BZ' },
\eeq{1.21q}
the expression for $\widetilde{Y}({\underline\Bu}'',{\underline\BG}'')$ can be written equivalently
as 
\beq \widetilde{Y}({\underline\Bu}'',{\underline\BG}'')=\int_\GG
\pmatrix{\sqcap {\underline\Bu}'' \cr -{\underline\CG}''}\cdot\BCL
\pmatrix{\sqcap {\underline\Bu}'' \cr -{\underline\CG}''}-2\Bh'\cdot{\underline\Bu}''.
\eeq{1.21r}

These two minimization variational principles are not the only ones one can derive. Indeed,
there is a continuous two-parameter family of related variational principles. The differential
constraints \eq{1.2} and the constitutive relation \eq{1.6} imply
\beq  \widetilde\CF=\sqcap \widetilde\Bu,\quad \widetilde\Bh+\sqcup\widetilde\CG=0,
\quad \widetilde\CG(\Bx)=\widetilde\BZ(\Bx)\widetilde\CF(\Bx), \eeq{1.22}
where
\beq \widetilde\Bu=e^{i\tau}\Bu,\quad \widetilde\CF=e^{i\tau}\CF, \quad 
\widetilde\CG=e^{i(\tau+\Gt)}\CG,\quad \widetilde\Bh=e^{i(\tau+\Gt)}\Bh,\quad
\widetilde\BZ(\Bx)=e^{i\Gt}\BZ(\Bx),
\eeq{1.23}
in which $\tau$ and $\Gt$ are real parameters.
So the fields with tildes on them
satisfy the same equations as those without tildes, and the variational principles
directly apply provided $\Gt$ is chosen so that
\beq \widetilde\BZ''(\Bx)\geq \Ga\BI,\quad\forall \Bx\in\GG,
\eeq{1.24}   
for some $\Ga>0$. 
With the subsequent replacement \eq{1.23} we obtain
a new set of variational principles parameterized by $\tau$ and $\Gt$.  
Observe that even if \eq{1.9a} is not satisfied for any $\Ga>0$ we still
may be able to find a range of values of $\Gt$ for which \eq{1.24}
holds, and thus for which we can obtain variational principles. With the
particular choice $\Gt=0$ and $e^{i\tau}=-i$ the variational principle
\eq{1.20} gets mapped to the dual variational principle \eq{1.21n}. 

All the previous minimization principles need the fundamental coercivity
assumption \eq{1.9a}. If $\BZ''(\Bx)=0$ in some part $\GY$ of the domain $\GG$
they cannot be applied directly. However we can still obtain variational
principles by taking appropriate limits.
If, in some region $\GY\subset\GG$, $\BZ''(\Bx)$ is very small in the sense that 
\beq \Gd\BI\geq\BZ''(\Bx),\quad\forall \Bx\in\GY,
\eeq{1.25}
for some very small $\Gd>0$, then within this region
\beq \pmatrix{\sqcap {\underline\Bu}' \cr -{\underline\CG}'}\cdot\BCL
\pmatrix{\sqcap {\underline\Bu}' \cr -{\underline\CG}'}
\approx [\BZ'\sqcap {\underline\Bu}'-{\underline\CG}']\cdot(\BZ'')^{-1}
[\BZ'\sqcap {\underline\Bu}'-{\underline\CG}'].
\eeq{1.26}
If $\GY=\GG$ then the variational principle reduces to a variational principle similar
to \eq{0.9a} which is not particularly interesting. If  $\GY\neq\GG$
the variational principle will only be useful if one takes
${\underline\CG}'\approx\BZ'\sqcap {\underline\Bu}'$ in $\GY$. The variational
principles still hold in the limit as $\Gd\to 0$ provided one first takes
\beq {\underline\CG}'=\BZ'\sqcap {\underline\Bu}',\quad\forall \Bx\in\GY,
\eeq{1.27}
corresponding to taking an exact solution of \eq{1.7} in this region (but with
$({\underline\Bu}',{\underline\BG}')$ not necessarily coinciding with the exact
solution $(\Bu',\BG')$ associated with the given boundary conditions). Then
as $\Gd\to 0$ the expression for $Y$ reduces to 
\beq
\widetilde{Y}({\underline\Bu}',{\underline\BG}')=\int_{\GG\setminus\GY}
\pmatrix{\sqcap {\underline\Bu}' \cr -{\underline\CG}'}\cdot\BCL
\pmatrix{\sqcap {\underline\Bu}' \cr -{\underline\CG}'}
-\int_{\GG}
2\Bh''\cdot{\underline\Bu}'.
\eeq{1.28}
Outside $\GY$ we still require that $\BZ''(\Bx)\geq \Ga\BI$ for some $\Ga>0$.

We now show how these variational principles may give useful information
about the Dirichlet to Neumann map on the boundary $\Md\GG$.
Let us assume $\Bh=0$ and introduce the $m$-component complex vector fields
\beq \Bq_0=\BG(\Bx)\cdot\Bn, \quad \Bu_0(\Bx)=\Bu(\Bx)\quad{\rm for}~\Bx\in\Md\GG.
\eeq{1.21h}
The relation between $\Bq_0$ and $\Bu_0$ must be linear and we can write 
$\Bq_0=\BN\Bu_0$ which defines the Dirichlet to Neumann map $\BN$. If
$\Bu_0^{(1)}$ and $\Bu_0^{(2)}$ denote two boundary conditions for $\Bu$ and
$\CG^{(j)}$, $\CF^{(j)}$, and $\Bq_0^{(j)}$, $j=1,2$, 
denote the associated fields then it follows from the key property and    
the symmetry of $\BZ(\Bx)$ that 
\beqa \langle \BN\Bu_0^{(1)},\Bu_0^{(2)}\rangle &= & 
\int_{\Md\GG}\Bq_0^{(1)}\cdot\Bu_0^{(2)} =  \int_{\GG}\CG^{(1)}\cdot\CF^{(2)} 
=\int_{\GG}\CF^{(1)}\cdot\BZ\CF^{(2)} \nonum
& = & \int_{\GG}\CG^{(2)}\cdot\CF^{(1)}
=\int_{\Md\GG}\Bq_0^{(2)}\cdot\Bu_0^{(1)}=\langle \Bu_0^{(1)},\BN\Bu_0^{(2)}\rangle,
\eeqa{1.21i}
which implies the map $\BN$ is symmetric. It has positive semidefinite imaginary part since
\beqa \langle \Bu_0',\BN''\Bu_0'\rangle & + & \langle \Bu_0'',\BN''\Bu_0''\rangle
 =  \int_{\Md\GG}\Bq_0''\cdot\Bu_0'-\Bq_0'\cdot\Bu_0''\nonum
& = & \int_{\GG}\CG''\cdot\CF'-\CG'\cdot\CF''
=\int_{\GG}\CF'\cdot\BZ''\CF'+\CF''\cdot\BZ''\CF''\geq 0.
\eeqa{1.21j}
Assuming $\BN''$ is invertible, we can write the relation $\Bq_0=\BN\Bu_0$ as 
\beq \pmatrix{\Bq_0'' \cr \Bu_0''}=\BCN\pmatrix{\Bu_0' \cr -\Bq_0'},
\eeq{1.21k}
where 
\beq \BCN=\pmatrix{\BN''+\BN'(\BN'')^{-1}\BN' & \BN'(\BN'')^{-1} \cr 
                   (\BN'')^{-1}\BN' & (\BN'')^{-1}}.
\eeq{1.21l}
Thus the map $\BCN$ giving the boundary fields $\Bq_0''$ and  $\Bu_0''$ in terms of
$\Bq_0'$ and  $\Bu_0'$ must have this special form. 

Now, using the key property, $Y(\Bu',\BG')$ can be expressed (when $\Bh=0$) in terms of the boundary
fields:
\beqa Y(\Bu',\BG') & = & 
\int_{\GG}\CG''\cdot\CF'-\CG'\cdot\CF''=\int_{\Md\GG}\Bq_0''\cdot\Bu_0'-\Bq_0'\cdot\Bu_0'' \nonum
& = & \langle\pmatrix{\Bu_0' \cr -\Bq_0'},\BCN\pmatrix{\Bu_0' \cr -\Bq_0'}\rangle,
\eeqa{1.21m}
and thus bounds on $Y(\Bu',\BG')$, obtained from the variational principle, give upper
bounds on the quadratic form associated with $\BCN$. Bounds on $\widetilde{Y}(\Bu'',\BG'')$,
obtained from the dual variational principle, give lower bounds on the quadratic form associated with $\BCN$.

\section{Application to Acoustics and Elastodynamics}
\setcounter{equation}{0}
Let us begin by studying the acoustic equation in $d$ dimensions:
\beq -\Div \BGr^{-1}\Grad P=(1/\Gk)\Go^2P, \eeq{2.1}
where $P(\Bx)$ is the complex pressure, $\Gk(\Bx)$ is the bulk modulus, $\BGr(\Bx)$ 
is the density tensor, and $\Go$ is the (fixed) frequency of oscillation.
[As is well known this equation with $d=2$ is also applicable in
cylindrical bodies, for antiplane elastodynamics where the displacement field
is directed parallel to the cylinder axis, and for electromagnetism 
when the fields are transverse electric (TE), where the electric field is
directed parallel to the cylinder axis, or transverse magnetic (TM), where the
magnetic field is directed parallel to the cylinder axis]. 
Normally one expects $\Gk$ to have a negative imaginary part (due to
bulk viscosity), and $\BGr$ to be $\Gr\BI$ where $\Gr(\Bx)$ is the positive
mass density. However, viewed as the effective density tensor of a composite material
at a given frequency $\Go$, 
$\BGr$ can be anisotropic (\citeAY{Schoenberg:1983:PPS})
and, at least in the context of elastodynamics, can
be complex valued with a positive semidefinite imaginary
part (\citeAY{Milton:2007:MNS}) with a real part that is not even necessarily positive
[as established by \citeAPY{Movchan:2004:SRR}, \citeAPY{Bouchitte:2004:HNR}, and \citeAPY{Felbacq:2005:TMM}
in the context of antiplane elastodynamics, where $1/\Gk$ plays the role of density,
and by  \citeAPY{Liu:2005:AMP} and \citeAPY{Avila:2005:BPI} in the context of
three-dimensional elastodynamics]. Such unusual effective densities are due
to the fact that different parts of the microstructure can experience different
accelerations, and can be out of phase with the overall applied force if this relative
motion dissipates energy.

Given the complex pressure field $P(\Bx)$ the associated complex velocity field of the fluid is
\beq \Bv=-i(\Go\BGr)^{-1}\Grad P. \eeq{2.1a}
[The associated physical pressure and physical velocity are respectively $(e^{-i\Go t}P(\Bx))'$ and 
$(e^{-i\Go t}\Bv(\Bx))'$ where the prime denotes the real part and $t$ denotes the time.]
Upon multiplying \eq{2.1} by $e^{i\Gt}$ and comparing it with \eq{1.7a}
we can make the identifications 
\beq \Bu=P,\quad \BL=-e^{i\Gt}\BGr^{-1},\quad \BK=0,\quad \BM=e^{i\Gt}\Go^2/\Gk,
\quad \Bh=0, \quad \BG=-ie^{i\Gt}\Go\Bv.
\eeq{2.2}
Thus the potential $\Bu$ is a scalar, corresponding to the case $m=1$. We 
choose $\Gt$ such that the imaginary part of
\beq \BZ=\pmatrix{-e^{i\Gt}\BGr^{-1} & 0 \cr 0 & e^{i\Gt}\Go^2/\Gk}
\eeq{2.3}
satisfies \eq{1.9a}. When $\BGr$ is real and positive definite, as is typically the
case, then we need to choose $\Gt$ in the range $0>\Gt>-\pi$. If in addition $\Gk$
is almost real and positive, with a small negative imaginary part, then we should
choose $\Gt$ appropriately small and negative. 

Explicit expressions for the
various variational principles in the acoustic case can of course be obtained
by making the substitutions \eq{2.2} in the relevant equations in the preceding section. 
For example, if $\BGr$ has a positive definite imaginary part (which is more applicable
in the mathematically analogous TE or TM electromagnetic problems) and $\Gk(\Bx)$ has
a negative imaginary part, then with $\Gt=0$ we have the variational principle
\beq Y(P',\Bv'')
=\inf_{{\underline P}'}\inf_{{\underline\Bv}''}Y({\underline P}',{\underline\Bv}''), 
\eeq{2.3a}
where 
\beq Y({\underline P}',{\underline\Bv}'')
=\int_{\GG}\pmatrix{\Grad{\underline P}' \cr -\Go{\underline\Bv}''}\cdot\BCR
\pmatrix{\Grad{\underline P}' \cr -\Go{\underline\Bv}''}+
\pmatrix{\Go{\underline P}' \cr -\Div{\underline\Bv}''}
\cdot\BCK
\pmatrix{\Go{\underline P}' \cr -\Div{\underline\Bv}''},
\eeq{2.3b}
in which
\beqa \BCR & = & \pmatrix{\Br''+\Br'(\Br'')^{-1}\Br' & \Br'(\Br'')^{-1} \cr 
                   (\Br'')^{-1}\Br' & (\Br'')^{-1}}, \nonum
\BCK & = & \pmatrix{k''+(k')^2/k'' & 
                   k'/k'' \cr 
                   k'/k'' & 1/k''},
\eeqa{2.3c}
and $\Br=-\BGr^{-1}$, and $k=1/\Gk$. In this variational principle the boundary values of 
${\underline P}'$ and ${\underline\Bv}''\cdot\Bn$ are fixed, and the infimums are
attained at fields $ P'$ and $\Bv''$ associated with solutions of the acoustic equation
satisfying the prescribed boundary conditions. That the Euler-Lagrange equations of
\eq{2.3a} coincide with \eq{2.1} and \eq{2.1a} follows from the analysis of \eq{1.21a}-\eq{1.21g},
and can also be shown directly.

When $\Br''$ is very small within
all of $\GG$ then we have
\beq \pmatrix{\Grad{\underline P}' \cr -\Go{\underline\Bv}''}\cdot\BCR
\pmatrix{\Grad{\underline P}' \cr -\Go{\underline\Bv}''}
\approx [\Br'\Grad{\underline P}'-\Go{\underline\Bv}'']\cdot(\Br'')^{-1}
 [\Br'\Grad{\underline P}'-\Go{\underline\Bv}''],
\eeq{2.3d}
and the variational principles will only be useful if we take trial fields
with $\Go{\underline\Bv}''\approx \Br'\Grad{\underline P}'$. The variational
principles still hold in the limit as $\Br''\to 0$ provided we first choose
${\underline\Bv}''=\Br'\Grad{\underline P}'/\Go$. Then, when $\BGr$ is real,
we have the variational principle
\beq Y( P')
=\inf_{{\underline P}'}Y({\underline P}')
\eeq{2.3e}
where
\beq  Y({\underline P}')=\int_{\GG}\pmatrix{\Go{\underline P}' \cr \Div\BGr^{-1}\Grad{\underline P}'/\Go}
\cdot\BCK
\pmatrix{\Go{\underline P}' \cr \Div\BGr^{-1}\Grad{\underline P}'/\Go}.
\eeq{2.3f} 
In this variational principle the boundary values of 
${\underline P}'$ and $\Bn\cdot\BGr^{-1}\Grad{\underline P}'$ are fixed, 
where the latter corresponds to fixing the boundary value of ${\underline\Bv}''\cdot\Bn$. 
The infimum is 
attained at a field $ P'$ associated with solutions of the acoustic equation
satisfying the prescribed boundary conditions. To show this, let $P'$ be a minimizer
and set ${\underline P}'=P'+S$ where $S=0$ and  $\Bn\cdot\BGr^{-1}\Grad S=0$ on $\Md\GG$.
A necessary condition for the minimum to be at $S=0$  is that the first derivative
of the functional $Y$ vanish: that is
\beq \int_{\GG} \pmatrix{U \cr \Go P''}\cdot\pmatrix{\Go S \cr \Div\BGr^{-1}\Grad S/\Go}=0,
\eeq{2.3fa}
where we have introduced
\beq \pmatrix{U \cr \Go P''}\equiv\BCK\pmatrix{\Go{\underline P}' \cr \Div\BGr^{-1}\Grad{\underline P}'/\Go}.
\eeq{2.3fb}
Integrating by parts the constraint \eq{2.3fa} gives
\beq \int_{\GG}S(\Go U+\Div\BGr^{-1}\Grad{\underline P}'')=0,
\eeq{2.3fc}
and this will be satisfied for any $S$ with $S=0$ and  $\Bn\cdot\BGr^{-1}\Grad S=0$ on $\Md\GG$
if and only if
\beq U=-\Div\BGr^{-1}\Grad{\underline P}''/\Go.
\eeq{2.3fd}
A straightforward calculation (similar to the equivalence of \eq{1.6} and \eq{1.15}) shows 
that \eq{2.3fb} is equivalent to \eq{2.1} with this value of $U$. Thus the 
acoustic equation is the Euler-Lagrange equation of \eq{2.3e}, when $P''$ is defined
by \eq{2.3fb}. This variational principle remains 
valid if we replace $P$ by $e^{i\Gj}P$ where $\Gj$ can be any constant. 

In a similar way, if $\BGr$ has a positive definite imaginary part but $\Gk$ is real
(which is more applicable to the analogous TM electromagnetic problem, where
the magnetic permeability plays the role of $1/\Gk$) then we have the variational principle
\beq Y(\Bv'')
=\inf_{{\underline \Bv}''}Y({\underline \Bv}''),
\eeq{2.3g}
where
\beq Y({\underline \Bv}'')=\int_{\GG}\pmatrix{\Grad(\Gk\Div{\underline\Bv}'')/\Go \cr -\Go{\underline\Bv}''}\cdot\BCR
\pmatrix{\Grad(\Gk\Div{\underline\Bv}'')/\Go \cr -\Go{\underline\Bv}''},
\eeq{2.3h}
in which the boundary values of  ${\underline\Bv}''\cdot\Bn$ and $\Gk\Div{\underline\Bv}''$
are fixed, where the latter corresponds to fixing the boundary value of ${\underline P}'$.

One has to slightly modify the analysis to apply it to the elastodynamic equations,
\beq -\Div(\BC\Grad\Bu)=\Bb+\Go^2\BGr\Bu,
\eeq{2.20}
in which $\Bu(\Bx)$ is the complex displacement field, $\Bb(\Bx)$ is the complex body-force density 
$\BGr(\Bx)$ is the complex density tensor, and $\BC(\Bx)$ is the complex elasticity tensor. Given 
the complex displacement field $\Bu(\Bx)$, the complex stress $\BGj(\Bx)$ and 
complex momentum density $p(\Bx)$ are given by
\beq \BGj=\BC\Grad\Bu,\quad \Bp=-i\Go\BGr\Bu. \eeq{2.20a}
[The associated physical displacement, body-force density, stress, and momentum density are 
respectively $(e^{-i\Go t}\Bu)'$, $(e^{-i\Go t}\Bb)'$, $(e^{-i\Go t}\BGj)'$ and $(e^{-i\Go t}\Bp)'$.]
Upon multiplying \eq{2.20} by $e^{i\Gt}$ and comparing it with \eq{1.7a}
we can make the identifications
\beq \BL=-e^{i\Gt}\BC,\quad \BK=0,\quad \BM=e^{i\Gt}\Go^2\BGr,\quad \Bh=e^{i\Gt}\Bb,
\eeq{2.21}
and \eq{2.20a} implies
\beq \BG=-e^{i\Gt}\BGj, \quad \Bg=ie^{i\Gt}\Go\Bp. \eeq{2.21a} 
The problem is that $\BC$ acting upon any antisymmetric matrix is zero,
and so $\BL$ and hence $\BZ$ are singular, which is a problem for computing
the inverses in the equation \eq{1.16} for $\BCL$. This problem is rectified in the
standard way, by replacing $\sqcap$ everywhere with $\widetilde\sqcap$ defined by
\beq \widetilde\sqcap \Bu\equiv\pmatrix{[\Grad \Bu+(\Grad\Bu)^T]/2 \cr \Bu}.
\eeq{2.22}
Then the fields $\CF$ and $\CG$ have the form \eq{1.1}, with $\BF$ and $\BG$
being symmetric $d\times d$ matrices: thus $\CF$ and $\CG$ can be represented
by $d(d+3)/2$ component vectors. We choose $\Gt$ such that the imaginary part of
\beq \BZ=\pmatrix{-e^{i\Gt}\BC & 0 \cr 0 & e^{i\Gt}\Go^2\BGr}
\eeq{2.23}
satisfies \eq{1.9a} on the space of these fields, and we compute the inverse
of $\BZ''$ on the space of these fields. When $\BGr$ is real and positive definite
then we need to choose $\Gt$ in the range $\pi>\Gt>0$. If in addition $\BC$
is almost real with a positive definite real part, and with a small negative definite imaginary part, then we should
choose $\Gt$ appropriately small and positive. With $\sqcap$ replaced by $\widetilde\sqcap$ the analysis of
equations \eq{1.14a}-\eq{1.14g} and \eq{1.21a}-\eq{1.21g} still holds, and we recover the elastodynamic equations
from the Euler-Lagrange equations associated with the variational principles.

In the special case $\Gt=0$ the functionals $Q$, $R$ and $Y$ reduce to
\beqa Q({\underline\Bu}',{\underline\Bu}'')
& = & \int_{\GG}(-{\underline\BGe}
\cdot\BC{\underline\BGe}+\Go^2{\underline\Bu}\cdot\BGr{\underline\Bu}
+2\Bb\cdot{\underline\Bu})'',
\nonum
R({\underline\BGj}',{\underline\BGj}'')
& = & \int_{\GG}[-{\underline\BGj}\cdot\BC^{-1}{\underline\BGj}
+(\Div{\underline\BGj}+\Bb)\cdot(\Go^2\BGr)^{-1}(\Div{\underline\BGj}+\Bb)]'',
\nonum
Y(({\underline\Bu}',{\underline\BGj}'))  =  \int_{\GG}
& - & \pmatrix{{\underline\BGe}' \cr -{\underline\BGj}'}\cdot\BCC
\pmatrix{{\underline\BGe}' \cr -{\underline\BGj}'} \nonum 
& + & \pmatrix{\Go{\underline\Bu}' \cr (\Bb'+\Div{\underline\BGj}')/\Go}\BCP
\pmatrix{\Go{\underline\Bu}' \cr (\Bb'+\Div{\underline\BGj}')/\Go}
+2\Bb''\cdot{\underline\Bu}',
\eeqa{2.24}
where
\beq
{\underline\BGe}=[\Grad {\underline\Bu}+(\Grad{\underline\Bu})^T]/2,
\eeq{2.25}
and
\beqa  \BCC & = & \pmatrix{\BC''+\BC'(\BC'')^{-1}\BC' & \BC'(\BC'')^{-1} \cr 
                   (\BC'')^{-1}\BC' & (\BC'')^{-1}}, \nonum
\BCP & = & \pmatrix{\BGr''+\BGr'(\BGr'')^{-1}\BGr' & 
                   \BGr'(\BGr'')^{-1} \cr 
                   (\BGr'')^{-1}\BGr' & (\BGr'')^{-1}}
\eeqa{2.26}
are respectively negative definite and positive definite matrices. 
It follows directly from the variational principles of \citeAPY{Gurtin:1964:VPL}
and \citeAPY{Willis:1981:VPDP},
applied to the time harmonic case considered here, that the stationary points
of the functionals $Q$ and $R$ correspond to solutions of the elastodynamics equations,
in agreement with the results of the previous section.

When $\BGr''$ is very small we have
\beq \pmatrix{\Go{\underline\Bu}' \cr (\Bb'+\Div{\underline\BGj}')/\Go}\BCP
\pmatrix{\Go{\underline\Bu}' \cr (\Bb'+\Div{\underline\BGj}')/\Go}
\approx 
[\Go^2\BGr'{\underline\Bu}'+\Bb'+\Div{\underline\BGj}'](\Go^2\BGr'')^{-1}
[\Go^2\BGr'{\underline\Bu}'+\Bb'+\Div{\underline\BGj}']
\eeq{2.27}  
and the variational principles only will be useful if we take trial fields
with $\Go^2\BGr'{\underline\Bu}'+\Bb'+\Div{\underline\BGj}'$ approximately
zero.  The variational principle involving the functional $Y$  still 
holds in the limit as $\BGr''\to 0$ provided we first choose
\beq
{\underline\Bu}'=-(\BGr')^{-1}(\Bb'+\Div{\underline\BGj}')/\Go^2.
\eeq{2.27a} 
Then, when $\BGr$ is real, we have the variational principle
\beq Y(\BGj')=\inf_{{\underline\BGj}'}Y({\underline\BGj}'),
\eeq{2.28}
where
\beq Y({\underline\BGj}')=\int_{\GG}-\pmatrix{{\underline\BGe}' \cr -{\underline\BGj}'}\cdot\BCC
\pmatrix{{\underline\BGe}' \cr -{\underline\BGj}'}
+2\Bb''\cdot{\underline\Bu}',
\eeq{2.29}
in which ${\underline\Bu}'$ and ${\underline\BGe}'$ are given by \eq{2.27a} and
\eq{2.25}. In this variational principle the trial field ${\underline\BGj}'$ can
be any real symmetric matrix valued field with prescribed values of 
${\underline\BGj}'\cdot\Bn$ and $\BGr^{-1}\Div{\underline\BGj}'$ at the boundary $\Md\GG$,
where prescribing $\BGr^{-1}\Div{\underline\BGj}'$ at $\Md\GG$ corresponds to prescribing
${\underline\Bu}'$ according to \eq{2.27a}. The infimum is 
attained at a field $\BGj'$ associated with solutions of the elastodynamic equation
satisfying the prescribed boundary conditions.

\section{The saddle point variational principles for electromagnetism}
\setcounter{equation}{0}
A comparison of the continuum elastodynamic equation \eq{2.20}
with Maxwell's equations, 
\beq \Curl\BE=i\Go\BB,\quad \Curl\BH=-i\Go\BD+\Bj,\quad 
\BD=\BGve\BE,\quad\BB=\BGm\BH,
\eeq{3.0} 
which can be rewritten (\citeAY{Milton:2006:CEM})
in the form
\beq -\Div(\BC\Grad\BE)=i\Go\Bj+\Go^2\BGve\BE,
 \eeq{3.1}
where now
\beq C_{ijk\ell}=e_{ijm}e_{k\ell n}\{\BGm^{-1}\}_{mn},
\eeq{3.2}
in which $\BE$ is the electric field, $\Bj$ is the free current
$\BGve$ the electric permittivity tensor,
$\BGm$ the magnetic permeability tensor, and $e_{ijm}=1$ (-1) if $(i,j,m)$
is an even (odd) permutation of $(1,2,3)$ and is zero otherwise, 
suggests that the preceeding analysis should
also extend to three-dimensional electromagnetism.  Upon
multiplying \eq{3.1} by $e^{i\Gt}$ and comparing it with \eq{1.7a}
we can make the identifications
\beq \Bu=\BE, \quad \BL=-e^{i\Gt}\BC,\quad \BK=0,\quad \BM=e^{i\Gt}\Go^2\BGve,
\quad \Bh=i\Go e^{i\Gt}\Bj.
\eeq{3.3}
The problem is now that $\BC$ acting upon any symmetric matrix is zero,
and so $\BL$ and hence $\BZ$ are singular, which is a problem for computing
the inverses in the equation \eq{1.16} for $\BCL$. This problem is rectified
by replacing $\sqcap$ everywhere with $\widehat\sqcap$ defined by
\beq \widehat\sqcap \BE\equiv\pmatrix{[\Grad \BE-(\Grad\BE)^T]/2 \cr \BE}. 
\eeq{3.4}
Then the fields $\CF$ and $\CG$ have the form \eq{1.1}, with $\BF$ and $\BG$
being antisymmetric $3\times 3$ matrices: thus $\CF$ and $\CG$ can be represented
by $6$ component vectors. We choose $\Gt$ such that the imaginary part of
\beq \BZ=\pmatrix{-e^{i\Gt}\BC & 0 \cr 0 & e^{i\Gt}\Go^2\BGve}
\eeq{3.5}
satisfies \eq{1.9a} on the space of these fields, and we compute the inverse
of $\BZ''$ on the space of these fields. 

Let us assume, for simplicity, that
\beq  \BGve''(\Bx)\geq \Ga_1\BI,\quad 
      \BGm''(\Bx)\geq \Ga_2\BI,\quad\forall \Bx\in\GG,
\eeq{3.6}
for some $\Ga_1>0$ and $\Ga_2>0$. (By multiplying \eq{3.1} by $e^{i\Gt}$
and redefining $\BGve$, $\BGm$ and $\Bj$ if necessary.) Then we can
take $\Gt=0$ and \eq{1.9a} holds for some $\Ga>0$. From \eq{3.0}
and \eq{3.4} we see that
\beq  \CF=\widehat\sqcap \BE=\pmatrix{\BF \cr \BE},\quad{\rm with~}
\BF=\frac{i\Go}{2}\pmatrix{0 & -B_3 & B_2 \cr
                           B_3 & 0 & -B_1 \cr
                           -B_2 & B_1 & 0},
\eeq{3.7}
and the constitutive law \eq{1.6ab} implies
\beq \CG=\pmatrix{\BG \cr \Go^2\BD} \quad{\rm with~}
\BG=-i\Go\pmatrix{0 & -H_3 & H_2 \cr
                           H_3 & 0 & -H_1 \cr
                           -H_2 & H_1 & 0}.
\eeq{3.8}
Thus the key property \eq{1.4} reduces to
\beq \int_{\GG}\Go^2(\BB\cdot\BH+\BE\cdot\BD)+i\Go\Bj\cdot\BE
=\int_{\Md\GG}-i\Go(\BH\times\Bn)\cdot\BE
=\int_{\Md\GG}-i\Go(\BE\times\BH)\cdot\Bn,
\eeq{3.9}
and holds for all fields $\BE$, $\BD$, $\BB$,and $\BH$ satisfying the differential
constraints
\beq \Curl\BE=i\Go\BB,\quad \Curl\BH=-i\Go\BD+\Bj. \eeq{3.10}
Now the first relation \eq{1.12} will hold if
\beq \int_{\Md\GG}(\BG''\cdot\Bn)\cdot\Bs=\int_{\Md\GG}-i\Go(\BH''\times\Bn)\cdot\Bs=0,
\eeq{3.11}
and for this it suffices that the tangential component of $\Bs$ vanishes at the
surface $\Md\GG$. Similarly the second relation in \eq{1.12} will hold if this 
condition is satisfied. Therefore the infimum and supremum in \eq{1.11} can be extended to
all ${\underline\Bu}'$ and ${\underline \Bu}''$ with the same tangential components as, respectively, 
$\BE'$ and $\BE''$. Similarly in the minimization variational principle \eq{1.20}, or \eq{1.21n}, we only need
require that ${\underline\Bu}'$, or ${\underline \Bu}''$, have the same tangential components as  
$\BE'$, or $\BE''$. 

With $\sqcap$ replaced by $\widehat\sqcap$ the analysis of
equations \eq{1.14a}-\eq{1.14g} and \eq{1.21a}-\eq{1.21g} still holds, and we recover Maxwell's equations
from the Euler-Lagrange equations associated with the variational principles.
If we introduce the tangential components     
\beq \BE_\|(\Bx)=\BE-\Bn(\Bn\cdot\BE), \quad \BH_\|(\Bx)=\BH-\Bn(\Bn\cdot\BH)
\eeq{3.11a}
of the fields $\BE(\Bx)$ and $\BH(\Bx)$ at the boundary $\Md\GG$, then 
\beq \Bq_0(\Bx)=\BG\cdot\Bn=-i\Go(\BH_\|\times\Bn).
\eeq{3.11b}
When $\Bj=0$ we can write $\Bq_0=\BN\BE_\|$, thus defining the map $\BN$ which governs the
electrodynamic response of the body at the frequency $\Go$. The analysis of \eq{1.21h}-\eq{1.21m} 
shows that the map $\BN$ is symmetric with positive semidefinite imaginary part, and can
be bounded (when $\Bj=0$) using the minimization variational principles.

Equivalently we can find the variational principles directly from Maxwell's
equations \eq{3.0}. The key property (basically Poynting's theorem)
\eq{3.9} is obtained by integrating
over $\GG$ the identity
\beq \Div(\BE\times\BH)=\BH\cdot(\Curl\BE)-\BE\cdot(\Curl\BH)
=i\Go(\BH\cdot\BB+\BE\cdot\BD)-\Bj\cdot\BE.
\eeq{3.12}
Motivated by the form of \eq{3.7} and \eq{3.8} let us redefine the fields
\beq \CF=\pmatrix{\BF \cr \Bf}=\pmatrix{i\Go\BB \cr \BE},\quad
\CG=\pmatrix{\BG \cr \Bg}=\pmatrix{-i\Go\BH \cr \Go^2\BD},
\eeq{3.13}
which are subject to the differential constraints that 
\beq \CF=\GO\BE,\quad i\Go\Bj+\mho\CG=0, \eeq{3.14}
where the operators $\GO$ and $\mho$ are defined by
\beq \GO\BE=\pmatrix{\Curl\BE \cr \BE},\quad \mho\CG=\Curl\BG+\Bg
\eeq{3.15}
so that the differential constraints \eq{3.10} and \eq{3.14} are equivalent.
The key property \eq{3.9} now takes the form
\beq \int_{\GG}\CG\cdot\CF+i\Go\Bj\cdot\BE
=\int_{\Md\GG}(\BE\times\BG)\cdot\Bn, \eeq{3.16}
where $\CG\cdot\CF\equiv\BF\cdot\BG+\Bf\cdot\Bg$.
The constitutive law $\CG=\BZ\CF$ holds with $\BZ$ redefined as
\beq \BZ=\pmatrix{-{\BGm}^{-1} & 0 \cr 0 & \Go^2\BGve}, \eeq{3.17}
and this can be reexpressed in the form \eq{1.9}, with 
\beq 
\CF'=\GO\BE'=\pmatrix{-\Go\BB'' \cr \BE'},\quad 
\CF''=\GO\BE''=\pmatrix{\Go\BB' \cr \BE''},\quad
\CG'=\pmatrix{\Go\BH'' \cr \Go^2\BD'},\quad \CG''=\pmatrix{-\Go\BH' \cr \Go^2\BD''},
\eeq{3.17a}
where these fields satisfy the differential constraints
\beq \CF'=\GO\BE',\quad -\Go\Bj''+\mho\CG'=0,\quad
\CF''=\GO\BE'',\quad
\Go\Bj'+\mho\CG''=0, 
\eeq{3.18}
implying, as a corollary of \eq{3.16}, that
\beq \int_{\GG}\CG'\cdot\GO\Bs-\Go\Bj''\cdot\Bs=0,\quad
 \int_{\GG}\CG''\cdot\GO\Bs+\Go\Bj'\cdot\Bs=0,
\eeq{3.19}
for all real valued vector fields $\Bs$ such that the tangential component of $\Bs$ vanishes at the
surface $\Md\GG$. Introducing the functional 
\beqa  Q({\underline\BE}',{\underline\BE}'') & = & \int_\GG 
\pmatrix{\GO {\underline\BE}' \cr \GO {\underline\BE}''}\cdot\pmatrix{\BZ'' & \BZ' \cr \BZ' & -\BZ''}
\pmatrix{\GO {\underline\BE}' \cr \GO {\underline\BE}''}+2\Go(\Bj'\cdot{\underline\BE}'-\Bj''\cdot{\underline\BE}'') \nonum
& = & \int_\GG [(\GO{\underline\BE})\cdot\BZ\GO {\underline\BE}+2i\Go\Bj\cdot\BE]'',
\eeqa{3.20}
we see from \eq{3.19} that
\beqa Q(\BE'+\Bs,\BE'')
&= & Q(\BE',\BE'')+\int_{\GG}(\GO\Bs)\cdot\BZ''\GO\Bs, \nonum
Q(\BE',\BE''+\Bs)& = & Q(\BE',\BE'')-
\int_{\GG}(\GO\Bs)\cdot\BZ''\GO\Bs.
\eeqa{3.21}
Consequently the pair of fields $({\underline\BE}',{\underline\BE}'')=(\BE',\BE'')$ is at the saddle point of \eq{1.11}
and of \eq{1.11a}, where the infimum and supremum are over all vector fields ${\underline\BE}'$ and ${\underline\BE}''$ 
with the same tangential components at the boundary $\Md\GG$ as, respectively, $\BE'$ and $\BE''$. Due to the diagonal
structure \eq{3.17} of the matrix $\BZ(\Bx)$ the formula \eq{3.20} for $Q({\underline\BE}',{\underline\BE}'')$ reduces
to
\beq Q({\underline\BE}',{\underline\BE}'') = \int_\GG
[(\Curl{\underline\BE})\cdot\Bm(\Curl{\underline\BE})+\Go^2\BE\cdot\BGve\BE+2i\Go\Bj\cdot\BE]'', 
\eeq{3.21aa}
where $\Bm=-\BGm^{-1}$, like $\BGve$, has positive definite imaginary part.  Hence we have the saddle point variational principles
\beq Q(\BE',\BE'')=
 \inf_{{\underline\BE}'}\sup_{{\underline\BE}''}Q({\underline\BE}',{\underline\BE}'')
=\sup_{{\underline\BE}''}\inf_{{\underline\BE}'}Q({\underline\BE}',{\underline\BE}''),
\eeq{3.21ab}
where
\beqa Q(\BE',\BE'') & = &\int_\GG \CG''\cdot\CF'+\CG'\cdot\CF''+2\Go(\Bj'\cdot{\underline\BE}'-\Bj''\cdot{\underline\BE}'')\nonum
& = & \int_\GG \Go^2(\BH'\cdot\BB''+\BH''\cdot\BB'+\BE'\cdot\BD''+\BE''\cdot\BD')+2\Go(\Bj'\cdot{\underline\BE}'-\Bj''\cdot{\underline\BE}'). \nonum &~&
\eeqa{3.21ac}
When $\Bj=0$ we can use \eq{3.12}
to express $Q(\BE',\BE'')$ just in terms of the tangential components of the fields at the boundary:
\beq Q(\BE',\BE'')= \Go\int_{\Md\GG}(\BE''\times\BH'')\cdot\Bn-(\BE'\times\BH')\cdot\Bn\quad{\rm when}~\Bj=0.
\eeq{3.21ad}

The result that the stationary point of the variational principle \eq{3.21ab} corresponds to a
solution of Maxwell's equations follows from a variational principle
derived by \citeAPY{Willis:1984:VPOE}. The Willis variational principle
is expressed in terms of the electromagnetic potentials 
$V(\Bx)$ and $\BA(\Bx)$, which are associated with $\BE$ and $\BB$ through the relations
\beq \BE=-\Grad V+i\Go\BA, \quad \BB=\Curl\BA.
\eeq{3.21ae}
and, in the time harmonic case, involves (to within a proportionality constant)
the functional
\beq \widehat Q({\underline V},{\underline \BA})=\Go^2
\int_{\Md\GG}\{[-(\Curl{\underline\BA})\cdot\Bm(\Curl{\underline\BA})+(\Grad {\underline V}
-i\Go{\underline\BA})\cdot\BGve(\Grad {\underline V}-i\Go{\underline\BA})+2i{\underline V}\Div\Bj/\Go-2{\underline\BA}\cdot\Bj]e^{-i\Go t}\}',
\eeq{3.21af}
where, as usual, the prime denotes the real part. His variational principle states that for any choice of $t$ the functional 
$\widehat Q({\underline V},{\underline \BA})$ 
applied to trial fields ${\underline V}$ and ${\underline \BA}$ having any given fixed
values at the boundary $\Md\GG$, has a stationary point at the potentials $V$ and $\BA$ solving
Maxwell's equations, and having the prescribed boundary values. 
Setting ${\underline \BE}= -\Grad {\underline V}+i\Go{\underline\BA}$
one sees by integrating by parts
that $\widehat Q({\underline V},{\underline \BA})$ is just a functional of ${\underline \BE}$ alone, and when $t$ is chosen
with $e^{-i\Go t}=-i$ one has the identity
\beq \widehat Q({\underline V},{\underline \BA})=Q({\underline\BE}',{\underline\BE}'').
\eeq{3.21ag}
Furthermore the values of ${\underline V}$ and ${\underline \BA}$ on $\Md\GG$ determine the tangential (but not normal) 
components of ${\underline\BE}'$ and ${\underline\BE}''$ on $\Md\GG$.

There is also a dual variational principle. By taking real and imaginary parts of
the constitutive relation $\CF=\BK\CG$ where $\BK=\BZ^{-1}$ has negative 
definite imaginary part we obtain \eq{1.14a}.
Let us introduce the functional
\beq R({\underline\BH}',{\underline\BH}'')=\int_\GG  
\pmatrix{{\underline\CG}' \cr {\underline\CG}''}\pmatrix{\BK'' & \BK' \cr \BK' & -\BK''}\pmatrix{{\underline\CG}' \cr {\underline\CG}''}
=\int_\GG(\underline\CG\cdot\BK\underline\CG)'',
\eeq{3.21a}
where ${\underline\CG}'$ and ${\underline\CG}''$ are given by
\beq {\underline\CG}'=\Go\pmatrix{{\underline\BH}'' \cr \Bj''-\Curl{\underline\BH}''}, \quad
 {\underline\CG}''=\Go\pmatrix{-{\underline\BH}' \cr -\Bj'+\Curl{\underline\BH}'},
\eeq{3.21b}
(to ensure that  $-\Go\Bj''+\mho{\underline\CG}'=0$ and  $\Go\Bj'+\mho{\underline\CG}''=0$ inside $\GG$) 
and ${\underline\BH}''$ and ${\underline\BH}'$ have the same tangential components as, respectively, 
$\BH''$ and $\BH'$. We then have the saddle point variational principles
\beq R(\BH',\BH'')=\sup_{{\underline\BH}''}\inf_{{\underline\BH}'}R({\underline\BH}',{\underline\BH}'') 
=\inf_{{\underline\BH}'}\sup_{{\underline\BH}''}R({\underline\BH}',{\underline\BH}''),
\eeq{3.21c}
where the supremum and infimum are over fields with the
required tangential components at the boundary.  Due to the diagonal
structure \eq{3.17} of $\BZ(\Bx)$ the formula for $R({\underline\BH}',{\underline\BH}'')$ reduces to
\beq
R({\underline\BH}',{\underline\BH}'') = 
\int_\GG [(\Bj-\Curl{\underline\BH})\cdot\Be(\Bj-\Curl{\underline\BH})+\Go^2\underline\BH\BGm\underline\BH]'',
\eeq{3.21d}
where $\Be=-\BGve^{-1}$, like $\BGm$, has positive definite imaginary part. From this expression one can
see that $R({\underline\BH}',{\underline\BH}'')$ is a special case of the functional 
associated with the \citeAPY{Willis:1984:VPOE} dual
variational principle. We also have that
\beq R(\BH',\BH'')=\int_\GG \CG''\cdot\CF'+\CG'\cdot\CF''=\int_\GG \Go^2(\BH'\cdot\BB''+\BH''\cdot\BB'+\BE'\cdot\BD''+\BE''\cdot\BD'),
\eeq{3.21e}
which when $\Bj=0$ can be expressed in terms of the tangential components of the fields at the boundary:
\beq  R(\BH',\BH'')
=\Go\int_{\Md\GG}(\BE''\times\BH'')\cdot\Bn
-(\BE'\times\BH')\cdot\Bn\quad{\rm when}~\Bj=0.
\eeq{3.21f}   

Consider a medium which is locally isotropic and such that $\BGve(\Bx)$ is purely imaginary
(which may be a good approximation at low frequencies in conductive media) 
while $\BGm(\Bx)$ is purely real (and so \eq{3.6} only holds in the limiting sense with
$\Ga_2=0$). We can write
\beq \BGve=i\BI/(\Go\Gr), \quad \Be=-i\Go\Gr\BI, \quad \BGm=\Gm\BI, \eeq{3.21g}
where the resistivity $\Gr(\Bx)$ and the permeability $\Gm(\Bx)$ are purely real and positive. Then
the expression \eq{3.21d} reduces to
\beq R({\underline\BH}',{\underline\BH}'') = \int_\GG \Go\Gr(|\Bj'-\Curl{\underline\BH}'|^2-|\Bj''-\Curl{\underline\BH}''|^2)
+2\Go^2\Gm{\underline\BH}'\cdot{\underline\BH}''
\eeq{3.21h}
and then when $\Bj=0$ the saddle point variational principle \eq{3.21c} corresponds to the \citeAPY{Borcea:1999:AAQ}  
variational principle. (Note that $\Bj$ has a different meaning in that paper.)
Now Maxwell's equations still hold if we make the replacements
\beq \BGve\to-i\BGve, \quad \BGm \to i\BGm,\quad \BE\to i\BE,\quad \BB\to i\BB, \quad \BD\to\BD, \quad \BH\to\BH,
\quad \Bj\to\Bj,
\eeq{3.21i}
(which is a special case of the transformation \eq{1.23}) and with these replacements applied to \eq{3.21g}
the expression \eq{3.21d} reduces to  
\beq R({\underline\BH}',{\underline\BH}'') =  \int_\GG -2\Go\Gr (\Bj'-\Curl{\underline\BH}')\cdot(\Bj''-\Curl{\underline\BH}'')
+\Go^2\Gm (|{\underline\BH}'|^2-|{\underline\BH}''|^2),
\eeq{3.21j}
and when $\Bj=0$ the saddle point variational principle \eq{3.21c}
then corresponds to the other Borcea variational principle. Multiplying
$\BGve$ and dividing $\BGm$ by $e^{i\Gt}$, where $0>\Gt>-\pi/2$ leads to
a continuous family of saddle point variational principles which interpolate the
two Borcea variational principles. 

\section{Minimization variational principles for electromagnetism}
\setcounter{equation}{0}
To obtain a minimization principle we rewrite the 
constitutive law in the form \eq{1.15} where
$\BCL$ is given by \eq{1.16}. Then we introduce the quadratic functional
\beq  Y({\underline\BE}',{\underline\BH}'')=\int_\GG 
\pmatrix{\GO{\underline\BE}' \cr -{\underline\CG}'}\cdot\BCL
\pmatrix{\GO{\underline\BE}' \cr -{\underline\CG}'}+2\Go\Bj'\cdot{\underline\BE}',
\eeq{3.22}
where ${\underline\CG}'$ is given by
\beq {\underline\CG}'=\Go\pmatrix{{\underline\BH}'' \cr \Bj''-\Curl{\underline\BH}''},
\eeq{3.23}
(to ensure that  $-\Go\Bj''+\mho{\underline\CG}'=0$ inside $\GG$) and ${\underline\BE}'$ and ${\underline\BH}''$ have
the same tangential components at the boundary $\Md\GG$ as, respectively, $\BE'$ and $\Go\BH''$.
Given two vector fields $\Bs$ and $\Bt$ with zero tangential components at the boundary $\Md\GG$
then \eq{3.19} and the key property \eq{3.16} imply 
\beq \int_{\GG}\CG''\cdot(\GO\Bs)+\Go\Bj'\cdot\Bs-\CT\cdot(\GO\Bu'')=0,
\quad {\rm where}~\CT=\Go\pmatrix{\Bt \cr -\Curl\Bt}.
\eeq{3.24}
As a result we deduce that
\beq  Y(\BE'+\Bs,\BH''+\Bt)= Y(\BE',\BH'')+
\int_\GG \pmatrix{\GO\Bs \cr -\CT}\cdot\BCL
\pmatrix{\GO\Bs \cr -\CT}.
\eeq{3.25}
Consequently, because $\CL$ is positive definite under the assumption \eq{3.6}, we the minimization
variational principle
\beq Y(\BE',\BH'')=\inf_{{\underline\BE}'}\inf_{{\underline\BH}''}Y({\underline\BE}',{\underline\BH}''),
\eeq{3.26}
where the infimums are over vector fields ${\underline\BE}'$ and ${\underline\BH}''$ with the same tangential 
components at the boundary $\Md\GG$ as, respectively, $\BE'$ and $\BH''$. Using \eq{1.17} the
minimum value of $Y({\underline\BE}',{\underline\BH}'')$ is
\beq Y(\BE',\BH'')
=\int_\GG \Go^2(\BH'\cdot\BB''-\BH''\cdot\BB'+\BE'\cdot\BD''-\BE''\cdot\BD')+
2\Go\Bj'\cdot{\underline\BE}'.
\eeq{3.26aa}
Now the power dissipated
into heat in the body $\GG$, averaged over a cycle of oscillation is
\beqa & ~& W(\BE,\BH) \nonum 
& = & \int_{\GG}
\left[\frac{\Go}{2\pi}\int_{0}^{2\pi}
(\BE e^{-i\Go t})'\cdot\frac{\Md(\BD e^{-i\Go t})'}{\Md t}+
(\BH e^{-i\Go t})'\cdot\frac{\Md(\BB e^{-i\Go t})'}{\Md t}
+(\BE e^{-i\Go t})'\cdot(\Bj e^{-i\Go t})'
~dt\right] \nonum
& = & \frac{1}{2}\int_{\GG}\Go(\BH'\cdot\BB''-\BH''\cdot\BB'+\BE'\cdot\BD''-\BE''\cdot\BD')
+\BE'\cdot\Bj'+\BE''\cdot\Bj'',
\eeqa{3.26ab}
which is proportional to \eq{3.26aa} when $\Bj=0$. So the minimum value of $Y$ 
has a physical interpretation when $\Bj=0$.

Due to the diagonal structure \eq{3.17} of $\BZ$ the formula \eq{3.22} reduces to
\beqa
Y({\underline\BE}',{\underline\BH}'')& = & \int_\GG 
\pmatrix{\Go{\underline\BE}' \cr \Curl{\underline\BH}''-\Bj''}\cdot\BCE
\pmatrix{\Go{\underline\BE}' \cr \Curl{\underline\BH}''-\Bj''}\nonum &~&
+\pmatrix{\Curl{\underline\BE}' \cr -\Go{\underline\BH}''}\cdot\BCM
\pmatrix{\Curl{\underline\BE}' \cr -\Go{\underline\BH}''} 
+2\Go\Bj'\cdot{\underline\BE}',
\eeqa{3.26a}
where $\BCE(\Bx)$ and $\BCM(\Bx)$ are the positive definite matrices
\beqa \BCE & = & \pmatrix{\BGve''+\BGve'(\BGve'')^{-1}\BGve' & \BGve'(\BGve'')^{-1} \cr 
                   (\BGve'')^{-1}\BGve' & (\BGve'')^{-1}}, \nonum
\BCM & = & \pmatrix{\Bm''+\Bm'(\Bm'')^{-1}\Bm' & 
                   \Bm'(\Bm'')^{-1} \cr 
                   (\Bm'')^{-1}\Bm' & (\Bm'')^{-1}}.
\eeqa{3.26b}
in which $\Bm=-\BGm^{-1}$. Also the constitutive law \eq{1.15} reduces to 
\beq \pmatrix{\BD'' \cr \BE''}=\BCE\pmatrix{\BE' \cr -\BD'}, \quad \quad
 \pmatrix{\BH' \cr -\BB'}=\BCM\pmatrix{\BB'' \cr \BH''}.
\eeq{3.26c}
So given any arbitrary prescribed values for the tangential components 
${\underline\BE}'$ and ${\underline\BH}''$ at the boundary $\Md\GG$ 
the minimum in the variational principle \eq{3.26} 
will satisfy the Euler-Lagrange equation and therefore will correspond to a solution of Maxwell equations. 
From the  minimizing fields $\BE'$ and $\BH''$ one can, according to \eq{3.26c}, determine
\beqa \BE''& = & (\BGve'')^{-1}\BGve'\BE'+(\BGve'')^{-1}(\Curl{\underline\BH}''-\Bj'')/\Go, \nonum
\BH' & = & \Bm'(\Bm'')^{-1}\BH''-[\Bm''+\Bm'(\Bm'')^{-1}\Bm'](\Curl{\underline\BE}')/\Go,
\eeqa{3.26d}
and in particular one can determine the tangential components of these fields at the boundary. 

It is often the case that $\BGm$ is real, 
while $\BGve$ has a positive definite imaginary
part. In this event we redefine $\BZ$, $\CG$ and $\Bj$ by multiplying them by
$e^{i\Gt}$. Again there is a continuous two-parameter family of related variational principles. We rewrite Maxwell's
equations \eq{3.0} in the form
\beq \Curl\widetilde\BE=i\Go\widetilde\BB,\quad 
\Curl\widetilde\BH=-i\Go\widetilde\BD+\widetilde\Bj,\quad 
\widetilde\BD=\widetilde\BGve\widetilde\BE,\quad\widetilde\BB=\widetilde\BGm\widetilde\BH,
\eeq{3.27}
where
\beqa \widetilde\BE& = & e^{-i\Go t_0}\BE,\quad \widetilde\BD=e^{i(\Gt-\Go t_0)}\BD, \quad
\widetilde\BH=e^{i(\Gt-\Go t_0)}\BH,\quad \widetilde\BB=e^{-i\Go t_0}\BB,\nonum
\widetilde\Bj& = & e^{i(\Gt-\Go t_0)}\Bj,\quad \widetilde\BGve=e^{i\Gt}\BGve, \quad
\widetilde\BGm=e^{-i\Gt}\BGm,
\eeqa{3.28}
in which the real constant $t_0$ (which may be viewed as a time interval)
can be chosen freely, while $\Gt$ must be chosen so that 
\beq  {\widetilde\BGve}''(\Bx)\geq \Ga_1\BI,\quad 
      {\widetilde\BGm}''(\Bx)\geq \Ga_2\BI,\quad\forall \Bx\in\GG,
\eeq{3.28a}
for some $\Ga_1>0$ and $\Ga_2>0$. Then the fields with the tildes on them
also satisfy Maxwell's equations and the variational principles directly apply to them.
With the subsequent replacement \eq{3.28} we obtain
a new set of variational principles parameterized by $\Gt$ and $t_0$. When $\Gt$ is non-zero
the minimum value of the minimization principles no longer has a physical interpretation
as the time averaged power dissipation in the body.

In particular by choosing
$\Gt=0$ and $t_0$ with $e^{-i\Go t_0}=-i$ we recover the dual variational principle that
\beqa \widetilde{Y}({\underline\BE}'',{\underline\BH}')& = & \int_\GG 
\pmatrix{\Go{\underline\BE}'' \cr \Bj'-\Curl{\underline\BH}'}\cdot\BCE
\pmatrix{\Go{\underline\BE}'' \cr \Bj'-\Curl{\underline\BH}'}\nonum &~&
+\pmatrix{\Curl{\underline\BE}'' \cr \Go{\underline\BH}'}\cdot\BCM
\pmatrix{\Curl{\underline\BE}'' \cr \Go{\underline\BH}'} 
+2\Go\Bj''\cdot{\underline\BE}'',
\eeqa{3.29}
is minimized at ${\underline\BE}''=\BE''$ and ${\underline\BH}'=\BH'$,
under the constraint that  ${\underline\BE}''$ and ${\underline\BH}'$
have the same tangential values as, respectively, $\BE''$ and $\BH'$.
A corollary is that
\beq W({\underline\BE},{\underline\BH})\equiv[Y({\underline\BE}',{\underline\BH}'')+
\widetilde{Y}({\underline\BE}'',{\underline\BH}')]/(4\Go)
\eeq{3.30}
is minimized at 
${\underline\BE}={\underline\BE}'+i{\underline\BE}''=\BE$ and 
${\underline\BH}={\underline\BH}'+i{\underline\BH}''=\BH$,
under the constraint that  ${\underline\BE}$ and ${\underline\BH}$
have the same tangential values as, respectively $\BE$ and $\BH$.
By making the substitutions \eq{3.28} in \eq{3.26aa}
it is easy to check that the minimum value of $W({\underline\BE},{\underline\BH})$ 
is the average
power dissipation $W(\BE,\BH)$ given by \eq{3.26ab}, even when $\Bj$ is non-zero.
An appealing feature of the expression (4.39) is that it remains invariant under
the replacement $\underline\BE\to e^{-i\Go t_0}\underline\BE$, 
$\underline\BH\to e^{-i\Go t_0}\underline\BH$ for any real value of $t_0$. However a disadvantage
is that the boundary conditions are overprescribed: if one imposes boundary conditions
on the tangential values of ${\underline\BE}$ and ${\underline\BH}$ that
are not compatible with the moduli $\BGve(\Bx)$ and $\BGm(\Bx)$ then the fields
which minimize $W({\underline\BE},{\underline\BH})$ will not satisfy Maxwell's
equations.

When $\BGm''$, or equivalently $\Bm''$, is very small we have
\beq \pmatrix{\Curl{\underline\BE}' \cr -\Go{\underline\BH}''}\cdot\BCM
\pmatrix{\Curl{\underline\BE}' \cr -\Go{\underline\BH}''} 
\approx [\Bm'\Curl{\underline\BE}'-\Go{\underline\BH}'']\cdot(\Bm'')^{-1}
[\Bm'\Curl{\underline\BE}'-\Go{\underline\BH}'']
\eeq{3.31}
and the variational principle is only useful if we take trial fields
with ${\underline\BH}''\approx \Bm'\Curl{\underline\BE}'/\Go$. 
If we take ${\underline\BH}''=\Bm'\Curl{\underline\BE}'/\Go$
we still obtain a useful variational principle in the
limit when $\BGm$ is real. From \eq{3.26} and \eq{3.26a} we have
\beq
Y(\BE')=\inf_{{\underline\BE}'}Y({\underline\BE}'),
\eeq{3.32}
where
\beq 
 Y({\underline\BE}') =  \int_\GG 
\pmatrix{\Go{\underline\BE}' \cr -\Curl\BGm^{-1}(\Curl{\underline\BE}')/\Go-\Bj''}\cdot\BCE
\pmatrix{\Go{\underline\BE}' \cr -\Curl\BGm^{-1}(\Curl{\underline\BE}')/\Go-\Bj''}
+2\Go\Bj'\cdot{\underline\BE}'.
\eeq{3.33}
In this variational principle the infimum is over vector fields ${\underline\BE}'$
with prescribed tangential components of ${\underline\BE}'$ 
and $\BGm^{-1}\Curl{\underline\BE}'$ at the boundary $\Md\GG$. The infimum is 
attained at a field $\BE'$ associated with the solution of Maxwell's equations
satisfying the prescribed boundary conditions.

\section{Application to Electromagnetic Tomography}
\setcounter{equation}{0}

Here we consider the tomography problem: how can one recover information about the
functions $\BGve(\Bx)$ and $\BGm(\Bx)$ given 
measurements of the tangential components $\BE_\|(\Bx)$ and $\BH_\|(\Bx)$
of the fields $\BE(\Bx)$ and $\BH(\Bx)$ at the boundary $\Md\GG$?
When it is known that the body $\GG$ consists of small inclusions in a matrix this
question has been answered by \citeAPY{Ammari:2001:AFP}: see also the
review by \citeAPY{Ammari:2004:RSI}. For other geometries a different approach is
needed and the variational principles provide this. 
First observe from \eq{3.12} and \eq{3.26ab} that the power
dissipation $W(\BE,\BH)$, only depends on the tangential values of these
fields at the boundary
\beq W(\BE,\BH)=-\frac{1}{2}\int_{\Md\GG}(\BE'_\|\times\BH'_\|)\cdot\Bn+(\BE''_\|\times\BH''_\|)\cdot\Bn,
\eeq{4.1}
and, as could be expected physically, is the time-averaged value of the flux of
the Poynting vector into the body $\GG$. Thus when $\Bj=0$, and trial fields
${\underline\BE}'$ and ${\underline\BH}''$ are chosen having the same tangential values at the
boundary as, respectively $\BE'$ and $\BH''$, the inequality
\beq W(\BE,\BH)\leq Y({\underline\BE}',{\underline\BH}'')/(2\Go) 
\eeq{4.2}
provides for a given $\BGve(\Bx)$ and $\BGm(\Bx)$ a bound on the possible tangential
components of the fields ${\underline\BE}''$ and ${\underline\BH}'$.
Conversely if these tangential components have been measured the inequality provides one
constraint on the possible values of $\BGve(\Bx)$ and $\BGm(\Bx)$. Additional
inequalities on  $\BGve(\Bx)$ and $\BGm(\Bx)$ may be obtained by chosing
different trial fields, by working with other minimization variational principles
in the family parameterized by $t_0$ and $\Gt$,  and 
by conducting a series of physical experiments with various different sets of values
of the tangential components of the fields  $\BE(\Bx)$ and $\BH(\Bx)$ at the boundary $\Md\GG$.
It is not necessary that the  moduli $\BGve(\Bx)$ and $\BGm(\Bx)$ remain the same for each physical
experiment provided we have some model for how they vary from experiment to experiment.
For example, the frequency $\Go$ could be different for the different experiments if we
know that the permittivity has the form $\BGve(\Bx)=\BGve_0+i\BGs_o/\Go$ (or some other known
parametric dependence on $\Go$). The objective is then to constrain the possible 
values of the real functions $\BGve_0(\Bx)$ and $\BGs_o(\Bx)$.

One criticism of this approach is that it only utilizes information (in the case $\Gt=0$)
about the time-averaged power dissipation in the body $\GG$. When $t=\Gt=0$ the measured values of
the functions $\BE''$ and $\BH'$ around the boundary $\Md\GG$ are not used except to calculate 
the single scalar quantity $W(\BE,\BH)$. This seems like a tremendous waste of information.
However some extra information from the measured fields can be incorporated
if one has measurements of the boundary value fields $\BE_\|^{(j)}(\Bx)$ and $\BH_\|^{(j)}(\Bx)$
for $n$ experiments indexed by the integer $j=1,2,\ldots,n$, with the moduli
$\BGve(\Bx)$ and $\BGm(\Bx)$ remaining the same for each experiment. Then one can infer the
boundary values $\BE_\|$ and $\BH_\|$ associated with a linear combination of these fields:
\beq \BE_\|=\sum_{j=1}^n\Gl_j\BE_\|^{(j)},\quad \BH_\|=\sum_{j=1}^n\Gl_j\BH_\|^{(j)},
\eeq{4.3}
where $\Gl_1, \Gl_2, \ldots \Gl_n$ are a set of complex constants. Letting $\BE$ and $\BH$ denote
the corresponding electric and magnetic fields inside $\GG$, it follows that $W(\BE,\BH)$ 
can be expressed as the sum
\beqa W(\BE,\BH) & = & \sum_{j=1}^n\sum_{k=1}^n(\Gl_j'\Gl_k'+\Gl_j''\Gl_k'')W_{jk}+(\Gl_j''\Gl_k'-\Gl_k''\Gl_j')S_{jk} \nonum
& = &
\frac{1}{2}\sum_{j=1}^n\sum_{k=1}^n(\Gl_j'\Gl_k'+\Gl_j''\Gl_k'')(W_{jk}+W_{kj})+(\Gl_j''\Gl_k'-\Gl_k''\Gl_j')(S_{jk}-S_{kj})
\eeqa{4.4}
involving the real quantities
\beqa W_{jk} & = & -\frac{1}{2}\int_{\Md\GG}({\BE'_\|}^{(j)}\times{\BH'_\|}^{(k)})\cdot\Bn+({\BE''_\|}^{(j)}\times{\BH''_\|}^{(k)})\cdot\Bn, \nonum
 S_{jk} & = & -\frac{1}{2}\int_{\Md\GG}({\BE'_\|}^{(j)}\times{\BH''_\|}^{(k)})\cdot\Bn-({\BE''_\|}^{(j)}\times{\BH'_\|}^{(k)})\cdot\Bn.
\eeqa{4.5}
which can be extracted from the experimental measurements of the boundary fields. Thus $W(\BE,\BH)$ 
depends not only on the $n$ diagonal elements, $W_{jj}$,  which physically represent the time-averaged power 
dissipation in each experiment, but also on the $n(n-1)$ elements $W_{jk}+W_{kj}$ and $S_{jk}-S_{kj}$,
with $j> k$, which have no such physical interpretation. It makes sense to choose trial fields
which are also linear combinations:
\beqa {\underline\BE}'=\sum_{j=1}^n(\Gl_j{{\underline\BE}}^{(j)})'=
\sum_{j=1}^n\Gl_j'{{\underline\BE}'}^{(j)}-\Gl_j''{{\underline\BE}''}^{(j)}, \nonum
{\underline\BH}''=\sum_{j=1}^n(\Gl_j{{\underline\BH}}^{(j)})''=
\sum_{j=1}^n\Gl_j'{{\underline\BH}''}^{(j)}+\Gl_j''{{\underline\BH}'}^{(j)}.
\eeqa{4.5aa}
Then $Y({\underline\BE}',{\underline\BH}'')/(2\Go) -W(\BE,\BH)$ must be a positive quadratic form in the
$2n$ real variables $\Gl_1', \Gl_1'',\Gl_2',\Gl_2'' \ldots \Gl_n',\Gl_n''$, and the positive semi-definiteness
of the associated $2n$ by $2n$ matrix will give constaints on the functions $\BGve(\Bx)$ and $\BGm(\Bx)$.

We can also use the variational principle \eq{3.32} for tomography purposes. When $\BGm$ is real and
$\Bj=0$ the
inequality
\beq  W(\BE,\BH)\leq Y({\underline\BE}')/(2\Go)  
\eeq{4.5a}
applies. Often one knows in advance that the material inside $\GG$ is non-magnetic, so that $\BGm=\BI$ inside $\GG$. Then
from knowledge of $W(\BE,\BH)$ \eq{4.5a} provides a constraint on the possible values of the electrical permittivity $\BGve$
for every suitable choice of the trial field ${\underline\BE}'$. On the other hand, suppose the medium inside $\GG$ was two
phase with an unknown phase boundary and with two different values of $\BGm$ inside $\GG$. Then the variational
principle \eq{4.5a} will be useless unless the choice of trial field is carefully correlated with the unknown phase
boundary, so that the tangential component of $\BGm^{-1}(\Curl{\underline\BE}')$ is continuous across the phase
boundary.

When $\Bj\ne 0$ then we have the inequality
\beq  W(\BE,\BH)\leq W({\underline\BE},{\underline\BH}), \eeq{4.6}
where $W({\underline\BE},{\underline\BH})$ is given by \eq{3.30}
and the trial fields ${\underline\BE}$ and ${\underline\BH}$ are required
to have the same tangential values as, respectively, $\BE$ and $\BH$.
Given measurements of the tangential values of  $\BE$ and $\BH$,
and trial fields ${\underline\BE}$ and ${\underline\BH}$ we expect that this inequality
should provide a useful constraint on the possible values of $\BGve(\Bx)$ and $\BGm(\Bx)$.

\section*{Acknowledgements}
The authors are grateful for support from 
the Universit\'e de Toulon et du Var, the University of Utah, and from the
National Science Foundation through grant DMS-070978.
The authors are grateful to Russell Richins for checking the 
manuscript, and to David Dobson for helpful discussions.

\bibliography{/u/ma/milton/tcbook,/u/ma/milton/newref}

\end{document}